\documentclass[journal]{IEEEtran}
\IEEEoverridecommandlockouts
\usepackage{lineno}
\usepackage{cite}
\usepackage{hyperref}
\usepackage{amsmath,amssymb,amsfonts}
\usepackage{amsmath}
\usepackage{amsthm}
\DeclareMathOperator*{\argmax}{argmax}
\usepackage{graphicx}
\usepackage{textcomp}
\usepackage{xcolor}
\usepackage{graphicx}
\usepackage{float}
\usepackage{subfigure}
\usepackage{amsmath}
\usepackage{amsfonts,amssymb}
\usepackage{mathrsfs}
\usepackage{mathtools}
\usepackage{algorithm}
\usepackage{algorithmicx}
\usepackage{algpseudocode}
\usepackage{bm}
\usepackage{multirow}
\usepackage{array}
\usepackage{amssymb}
\usepackage{amsmath}
\usepackage{cite}
\usepackage{url}
\usepackage{xcolor}
\usepackage{cite,graphicx,amsmath,amssymb}
\usepackage{subfigure}
\usepackage{fancyhdr}
\usepackage{mdwmath}
\usepackage{mdwtab}
\usepackage{caption}
\usepackage{amsthm}
\usepackage{setspace}
\usepackage{bm}
\usepackage{algorithm}
\usepackage{algpseudocode}
\usepackage{mathtools}
\usepackage{dsfont}
\usepackage{bbm}
\usepackage{cases}
\newtheorem{remark}{Remark}
\theoremstyle{definition}
\newtheorem{theorem}{Theorem}

\newtheorem{lemma}{Lemma}

\newtheorem{corollary}{Corollary}

\makeatletter
\newcommand{\biggg}{\bBigg@{3}}
\newcommand{\Biggg}{\bBigg@{3.5}}
\makeatother
\def\BibTeX{{\rm B\kern-.05em{\sc i\kern-.025em b}\kern-.08em
		T\kern-.1667em\lower.7ex\hbox{E}\kern-.125emX}}
\expandafter\def\expandafter\normalsize\expandafter{%
	\normalsize%
	\setlength\abovedisplayskip{5pt}%
	\setlength\belowdisplayskip{5pt}%
	\setlength\abovedisplayshortskip{3pt}%
	\setlength\belowdisplayshortskip{3pt}%
}
\begin{document}
\title{On the Performance of Physical Layer Security for Continuous-Aperture Array (CAPA) Systems}
\author{Boqun~Zhao,~\IEEEmembership{Graduate Student Member,~IEEE,}~Chongjun~Ouyang,~\IEEEmembership{Member,~IEEE,}\\Xingqi~Zhang,~\IEEEmembership{Senior Member,~IEEE,}
	and Yuanwei~Liu,~\IEEEmembership{Fellow,~IEEE}\vspace{-13pt}
\thanks{B. Zhao and X. Zhang are with the Department of Electrical and Computer Engineering, University of Alberta, Edmonton AB, T6G 2R3, Canada (email: \{boqun1, xingqi.zhang\}@ualberta.ca).}
\thanks{C. Ouyang is with the School of Electronic Engineering and Computer Science, Queen Mary University of London, London E1 4NS, U.K. (e-mail: c.ouyang@qmul.ac.uk).}
\thanks{Y. Liu is with the Department of Electrical and Electronic Engineering, The University of Hong Kong, Hong Kong (email: yuanwei@hku.hk).}
}

\maketitle

\begin{abstract}
A continuous-aperture array (CAPA)-based secure transmission framework is proposed to enhance physical layer security. Continuous current distributions, or beamformers, are designed to maximize the secrecy transmission rate under a power constraint and to minimize the required transmission power for achieving a specific target secrecy rate. On this basis, the fundamental secrecy performance limits achieved by CAPAs are analyzed by deriving closed-form expressions for the maximum secrecy rate (MSR) and minimum required power (MRP), along with the corresponding optimal current distributions. To provide further insights, asymptotic analyses are performed for the MSR and MRP, which reveals that \romannumeral1) for the MSR, the optimal current distribution simplifies to maximal ratio transmission (MRT) beamforming in the low-SNR regime and to zero-forcing (ZF) beamforming in the high-SNR regime; \romannumeral2) for the MRP, the optimal current distribution simplifies to ZF beamforming in the high-SNR regime. The derived results are specialized to the typical array structures, e.g., planar CAPAs and planar spatially discrete arrays (SPDAs). The rate and power scaling laws are further analyzed by assuming an infinitely large CAPA. Numerical results demonstrate that: \romannumeral1) the proposed secure continuous beamforming design outperforms MRT and ZF beamforming in terms of both achievable secrecy rate and power efficiency; \romannumeral2) CAPAs achieve superior secrecy performance compared to conventional SPDAs.
\end{abstract}
\vspace{-3pt}

\section{Introduction}
Multiple-antenna technology is a fundamental pillar of modern wireless communication systems. Its core principle is to utilize a larger number of antenna elements in order to enhance spatial degrees of freedom (DoFs) and improve channel capacity \cite{tse2005fundamentals}. 

{Traditionally, multiple-antenna systems are designed with a spatially discrete topology, where each antenna is represented as an individual point in space. Driven by the benefits of integrating more antennas, the concept of densely packed antenna arrays has gained significant attention in the field of communications. By reducing the spacing between elements within a fixed array aperture, more antennas can be accommodated, thereby enhancing spatial DoFs. This progression has given rise to many state-of-the-art array architectures, such as massive or holographic multiple-input multiple-output (MIMO)  \cite{holographic,holographic2}, large intelligent surface \cite{intelligent_surface}, dynamic metasurface antennas \cite{meta}. In these systems, antenna elements are arranged in an ultra-dense configuration with spacings of less than half a wavelength, leading to improved spectral efficiency.

The holy grail of multiple-antenna systems is envisioned as the development of \emph{spatially continuous} electromagnetic (EM) apertures, referred to as \emph{continuous-aperture arrays (CAPAs)} \cite{liu2024capa}. A CAPA operates as a single electrically large-aperture antenna with a continuous current distribution, which comprises a (virtually) infinite number of radiating elements coupled with electronic circuits and a limited number of radio-frequency (RF) chains. On one hand, CAPAs fully utilize the entire aperture surface, which enables significant enhancements in spatial DoFs and array gains. On the other hand, they provide precise control over the amplitude and phase of the current across the aperture's surface. In summary, CAPAs can leverage spatial resources far more \emph{effectively} and \emph{flexibly} than traditional spatially discrete arrays (SPDAs). This capability allows them to approach the theoretical limits of channel capacity, positioning CAPAs as a cornerstone of next-generation wireless communications \cite{liu2024capa}.}

{ 
	
	However, it is important to emphasize that synthesizing and controlling continuous current distributions over CAPAs requires fully analog beamforming mechanisms at each aperture point \cite{BJORNSON20193,liu2024capa}, which are more sophisticated than the element-wise excitation in SPDAs. In particular, CAPAs typically involve surface wave generation, analog amplitude/phase modulation via metamaterials or photodetectors, and integrated control architectures \cite{liu2024capa}. Therefore, compared to the conventional SPDAs, where individual antenna elements are driven by discrete RF chains and their radiation patterns are fixed once the antenna type is determined, CAPA hardware may pose greater complexity in design, calibration, and real-time reconfigurability. Nevertheless, in recent years, rapid advancements in electromagnetic and optical materials, as well as in array fabrication techniques, have enabled the realization of functional CAPA prototypes \cite{smith2017analysis,araghi2021holographic,prather2017optically,yuan2024interdigital}. Some prototypes have even reached commercialization and, according to initial test reports, demonstrate substantial potential for improving coverage and throughput in practical wireless communication environments \cite{staff2019holographic,sazegar2022full}.
		
}

{Unlike traditional arrays, where the channel is typically modeled as a discrete superposition of array responses from a finite number of antenna elements, the CAPA channel is described by a continuous EM response between apertures' points. In particular, while the channels of conventional SPDAs are described using finite-dimensional matrices, the spatial response of a CAPA is modeled as a continuous operator in Hilbert space with infinite dimensions \cite{migliore2008electromagnetics,capa_single_0}. This discrete-to-continuous transition is not merely a technical adjustment but a paradigm shift in the conceptualization and design of wireless transmission systems \cite{liu2024capa}. It fundamentally alters the analytical and design frameworks and renders conventional methods developed for SPDAs unsuitable for CAPAs. Therefore, this shift necessitates the development of novel conceptual and mathematical tools tailored to address the continuous EM field interactions in CAPAs.}

\vspace{-5pt}
\subsection{Prior Works}
Recently, there has been growing research interest in the design and analysis of CAPA-based wireless communications. In \cite{capa_single_4}, the authors proposed a wavenumber-division multiplexing framework to enable multi-stream data transmission between two linear CAPAs. This concept was extended to CAPA-based multiuser MIMO channels in \cite{zhang2023pattern}, where a Fourier-based method was developed to maximize the downlink sum-rate by optimizing the current distribution used for modulating RF signals. Building on this approach, \cite{optimization} further studied beamforming design for uplink multiuser CAPA systems. More recently, \cite{wang2024beamforming} and \cite{wang2024optimal} proposed two calculus of variations (CoV)-based approaches for beamforming in unpolarized CAPA-based multiuser channels. Additionally, \cite{guo2024multi} applied deep learning techniques to design current distributions for multiuser CAPA systems.

In addition to continuous beamforming design, the performance of CAPA-based wireless communication systems has also been analyzed. In \cite{capacity}, the channel capacity between two spherical dielectric CAPAs was studied. The authors of \cite{xie} discussed the effective DoFs and capacity between two linear CAPAs. In \cite{wan_2}, the Fredholm determinant was utilized to compare the mutual information of CAPA- and SPDA-based MIMO channels, and this analysis was further extended in \cite{wan_1} to incorporate the effects of non-white EM interference. Moreover, \cite{zhao2024continuous} analyzed the sum-rate capacity and capacity region of CAPA-based multiuser uplink and downlink channels. Extensions to CAPA-based fading channels were presented in \cite{ouyang2024diversity} with a focus on the diversity-multiplexing tradeoff in the high signal-to-noise (SNR) region. Additionally, \cite{yindi} evaluated the signal-to-interference-plus-noise ratio in uplink CAPA systems and proposed an adaptive interference mitigation method. Building on these advancements, a recent study derived the optimal linear receive beamformer for CAPA-based multiuser channels and analyzed the achieved performance in terms of both sum-rate and mean-squared error (MSE) \cite{ouyang2024performance}. 

\vspace{-5pt}
\subsection{Motivation and Contributions}
The aforementioned works demonstrate the superiority of CAPAs over SPDAs in enhancing wireless communication performance. However, these studies mainly focus on analyzing or optimizing system effectiveness, such as sum-rate \cite{zhang2023pattern,optimization,wang2024beamforming,wang2024optimal,guo2024multi}, or system reliability, such as outage probability \cite{ouyang2024diversity} and MSE \cite{ouyang2024performance}. Beyond these metrics, another critical issue for wireless communication systems is their security. Specifically, the broadcast nature of wireless channels exposes transmitted signals to potentially insecure environments, making them susceptible to interception by eavesdroppers \cite{chen2016survey}. This vulnerability emphasizes the crucial need for ensuring robust wireless security \cite{chen2016survey}.

In the context of wireless security, secure channel coding has been theoretically proven as an effective approach to achieving nearly 100\% secure transmission at the physical layer \cite{chen2016survey}. This strategy, known as \emph{physical layer security (PLS)}, addresses the limitations of traditional cryptographic methods applied at higher layers (such as the network layer) by eliminating the need for additional spectral resources and reducing signaling overhead \cite{yang2015safeguarding}. The fundamental model for safeguarding information at the physical layer is Wyner's wiretap channel \cite{wyner}, which introduced the concept of secrecy transmission rate and secrecy channel capacity---the supremum of the achievable secure coding rate. In recent years, beamforming has been widely recognized as an effective method for enhancing secrecy capacity, thereby improving the PLS of wireless systems \cite{yang2015safeguarding}. Given that CAPAs are considered a promising architecture for achieving high beamforming gains, their application to PLS is particularly compelling. However, the use of CAPAs to enhance the performance for PLS remains underexplored in existing literature, which motivates this work.

This article aims to analyze the fundamental performance limits achieved by CAPA-based PLS. {Specifically, by proposing a novel approach that solves generalized Rayleigh quotient problems within continuous Hilbert spaces, we derive closed-form expressions for two key metrics: the \emph{maximum secrecy rate (MSR)} under a given power constraint and the \emph{minimum required power (MRP)} to achieve a specified secrecy rate target.} The main contributions of this work are summarized as follows.
\begin{itemize}
	\item We propose a CAPA-based transmission framework to enable secure communications with a legitimate user in the presence of an eavesdropper. Leveraging the EM theory and information theory, we introduce continuous operator-based signal and channel models to characterize CAPA-based secure communications. Within this framework, we define two critical metrics to evaluate secrecy performance: the MSR under a given power constraint and the MRP to achieve a specified secrecy rate target.
	\item For the problem of maximizing the secrecy rate, we derive a closed-form expression for the MSR and the optimal current distribution to achieve it. Additionally, we provide high-SNR and low-SNR approximations of the MSR and prove that the optimal current distribution simplifies to maximal ratio transmission (MRT) beamforming in the low-SNR regime and zero-forcing (ZF) beamforming in the high-SNR regime. Furthermore, we analyze the MSR achieved by a planar CAPA under a line-of-sight (LoS) model. To gain deeper insights, we conduct an asymptotic analysis of the achieved MSR by extending the aperture size to infinity and demonstrate that the asymptotic MSR adheres to the principle of energy conservation.
	\item For the problem of minimizing the required power to guarantee a target secrecy rate, we derive closed-form expressions for both the optimal current distribution and the MRP. On this basis, we demonstrate that the MRP outperforms that achieved by MRT beamforming and prove that the optimal current distribution simplifies to ZF beamforming when the target secrecy rate becomes infinitely large. Additionally, we derive a closed-form expression for the MRP achieved by a planar CAPA and characterize its asymptotic behavior as the aperture size approaches infinity.
	\item We provide numerical results to validate the effectiveness of CAPAs in enhancing PLS and demonstrate that: \romannumeral1) increasing the aperture size improves the secrecy performance achieved by CAPA-based secure beamforming; \romannumeral2) both the MSR and MRP converge to a finite constant as the aperture size of CAPA approaches infinity, which aligns with the principle of energy conservation; \romannumeral3) the proposed optimal continuous beamformers outperform the MRT and ZF-based schemes in terms of both MSR and MRP; and \romannumeral4) CAPA yields superior secrecy performance compared to conventional SPDAs.
\end{itemize}

The remainder of this paper is organized as follows. Section \ref{Section: System Model} introduces the CAPA-based secure transmission framework and defines the metrics of MSR and MRP. Sections \ref{sec_MSR} and \ref{sec_MRP} analyze the fundamental performance limits of the MSR and MRP achieved by CAPAs, respectively. Section \ref{numerical} presents numerical results to validate the theoretical findings. Finally, Section \ref{conclusion} concludes the paper.

\begin{figure}[!t]
	\centering
	\includegraphics[height=0.23\textwidth]{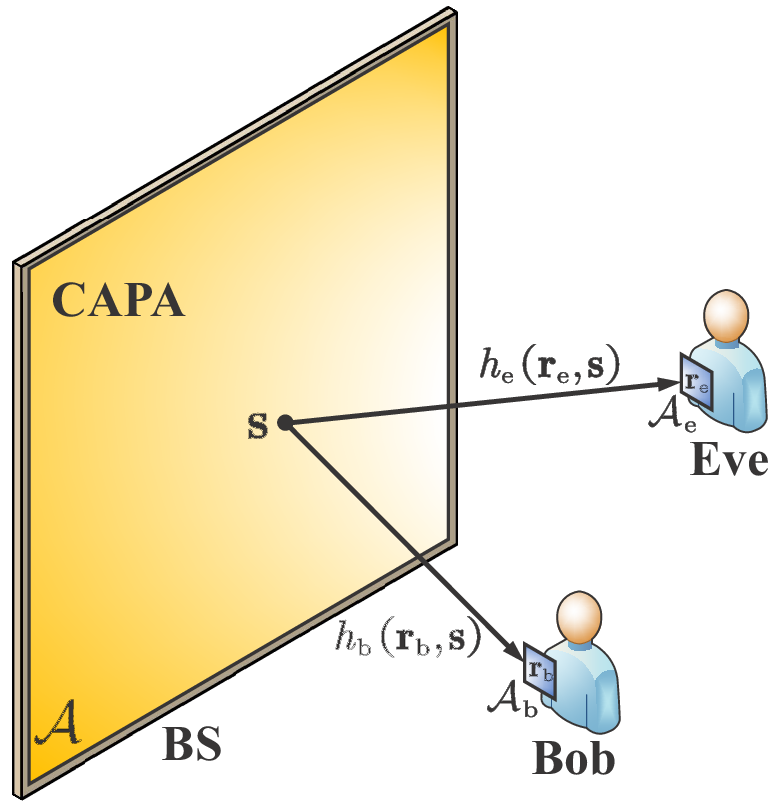}
	\caption{Illustration of a CAPA-based wiretap channel.}
	\vspace{-5pt}
	\label{Figure: System_Model}
\end{figure}

\section{System Model}\label{Section: System Model}
As illustrated in {\figurename} {\ref{Figure: System_Model}}, we consider a wiretap channel consisting of a base station (BS), a legitimate user (Bob, denoted as $\rm{b}$), and an eavesdropper (Eve, denoted as $\rm{e}$). Each entity is equipped with a CAPA. Let $\mathcal{A}\subseteq{\mathbb{R}}^{3\times1}$ represent the aperture of the CAPA at the BS, with its size given by $\lvert \mathcal{A} \rvert=\int_{\mathcal{A}}{\rm{d}}{\mathbf{s}}$. Moreover, let ${\mathcal{A}}_{k}\subseteq{\mathbb{R}}^{3\times1}$ denote the aperture of the CAPA at each user $k\in\{\rm{b},\rm{e}\}$, with the location $\mathbf{r}_k$ and aperture size $\lvert \mathcal{A}_k \rvert=\int_{{\mathcal{A}}_k}{\rm{d}}{\mathbf{r}}$. Both Bob and Eve are assumed to have complete channel state information (CSI) regarding their respective effective channels. {It is further assumed that Eve is a registered user, and thus, the BS obtains the CSI for both Eve and Bob during the channel training phase. Under these conditions, Eve is expected to receive the common messages broadcast across the network but remain uninformed about the confidential messages intended solely for Bob \cite{ly2010mimo}.}
\vspace{-5pt}
\subsection{CAPA-Based Transmission}\label{Section: System Model: CAPA-Based Transmission}
The BS intends to transmit a \emph{confidential message} $W$ to Bob over $N$ channel uses, and ensures that it remains secret from Eve. To achieve this, the BS first encodes the confidential message $W$ into the codeword $[c(1),\ldots,c(N)]$ using a properly designed encoder \cite{leung1978gaussian}. The encoded symbols are Gaussian distributed with zero mean and unit variance, i.e., $c(n)\sim{\mathcal{CN}}(0,1)$ for $n\in{\mathcal{N}}\triangleq\{1,\ldots,N\}$. 

{Next, the BS maps the encoded symbols over the time interval $n\in{\mathcal{N}}$, i.e., $c(n)$, into the transmit signal $x(\mathbf{s},n)\in{\mathbb{C}}$ by utilizing a source current $j({\mathbf{s}})$\footnote{{The source current $j({\mathbf{s}})$ can be interpreted as a continuous-version beamformer that encapsulates both amplitude and phase control at point $\mathbf{s}$. Implementing such a continuously defined source current that can achieve analog modulation of both amplitude and phase at each aperture point requires advanced hardware platforms. Possible realization involves metasurface-enabled leaky-wave antennas \cite{smith2017analysis,araghi2021holographic}, optically driven tightly coupled arrays \cite{prather2017optically}, and interdigital transducer-based grating antennas \cite{yuan2024interdigital}. Since this work focuses on the theoretical performance limit, detailed descriptions of these implementations are beyond the scope of this paper, while a comprehensive introduction can be found in \cite[Section II]{liu2024capa}.}}, where $\mathbf{s}\in\mathcal{A}$ denotes a spatial point on the BS aperture.} This signal is radiated towards Bob, while also being overheard by Eve. The transmit signal at the $n$th time interval is given by
\begin{equation}\label{Transmit_Signal}
{x}(\mathbf{s},n)=j({\mathbf{s}})c(n),~n\in{\mathcal{N}},
\end{equation}
where the source current is subject to the power constraint $\int_{\mathcal{A}}\lvert j({\mathbf{s}})\rvert^2
{\rm{d}}{\mathbf{s}}\leq P$. Therefore, the electric field excited by ${x}(\mathbf{s},n)$ at point $\mathbf{r}\in\mathcal{A}_k$ can be written as follows \cite{wang2024beamforming,wang2024optimal}:
\begin{align}\label{Electric_Field_Model}
e_k(\mathbf{r},n)&=\int_{{\mathcal{A}}}h_k(\mathbf{r},{\mathbf{s}}){x}({\mathbf{s}},n){\rm{d}}{\mathbf{s}}\notag\\
&=c(n)\int_{{\mathcal{A}}}h_k(\mathbf{r},{\mathbf{s}})j({\mathbf{s}})
{\rm{d}}{\mathbf{s}},
\end{align}
where $h_k(\mathbf{r},{\mathbf{s}})$ denotes user $k$'s spatial response from $\mathbf{s}$ to $\mathbf{r}$\footnote{{ $h_k(\mathbf{r},{\mathbf{s}})$ is based on electromagnetic propagation principles, and does not require independent estimation for each spatial pair $(\mathbf{s}, \mathbf{r})$. In LoS scenarios, the channel is a deterministic function of geometry, allowing the BS to compute $h_k(\mathbf{r},{\mathbf{s}})$ analytically once the user’s location is known. In multipath environments, by leveraging geometry-based models e.g., ray tracing \cite{jiang2024cram}, and discretization-based approach, e.g., Fourier series expansion \cite{capa_single_4,zhang2023pattern,optimization}, the BS can recover the full spatial channel using a limited set of parameters, thereby controlling the estimation overhead.}}. 

The total observation of user $k$ at point $\mathbf{r}\in{\mathcal{A}}_k$ is the sum of the information-carrying electric field $e_k(\mathbf{r},n)$ and a random noise field $z_k(\mathbf{r},n)$, i.e.,
\begin{align}\label{Total_Electric_Field_Model}
y_k(\mathbf{r},n)&=e_k(\mathbf{r},n)+z_k(\mathbf{r},n)\notag\\
&=c(n)\int_{{\mathcal{A}}}h_k(\mathbf{r},{\mathbf{s}})j({\mathbf{s}})
{\rm{d}}{\mathbf{s}}+z_k(\mathbf{r},n),
\end{align}
where $z_k(\mathbf{r},n)$ denotes the $n$th sample of additive white Gaussian noise (AWGN) at ${\mathbf{r}}$ with mean zero and variance (noise power) $\sigma_k^2$, i.e., $z_k(\mathbf{r},n)\sim {\mathcal{CN}}(0,\sigma_k^2)$. The SNR of user $k$ for decoding $c(n)$ is given by \cite{zhao2024continuous}
\begin{align}\label{Received_SNR_Basic}
	\gamma_k&=\frac{{\mathbb{E}}\{\lvert c(n)\rvert^2\}\left\lvert\int_{\mathcal{A}_k}\int_{{\mathcal{A}}}h_k(\mathbf{r},{\mathbf{s}})j({\mathbf{s}})
		{\rm{d}}{\mathbf{s}}{\rm{d}}{\mathbf{r}}\right\rvert^2}{\int_{\mathcal{A}_k}{\mathbb{E}}\{\lvert z_k(\mathbf{r},n)\rvert^2\}{\rm{d}}{\mathbf{r}}}\notag\\
		&=\frac{1}{\sigma_k^2 \lvert \mathcal{A}_k \rvert}{\left\lvert\int_{\mathcal{A}_k}\int_{{\mathcal{A}}}h_k(\mathbf{r},{\mathbf{s}})j({\mathbf{s}})
			{\rm{d}}{\mathbf{s}}{\rm{d}}{\mathbf{r}}\right\rvert^2}.
\end{align}
Given that the aperture size of each user is typically on the order of the wavelength, it is significantly smaller than both the propagation distance and the size of the BS aperture. Therefore, the variations in the channel response across the receive aperture are negligible, which yields
\begin{equation}
	h_k(\mathbf{r},{\mathbf{s}})\approx h_k(\mathbf{r}_k,{\mathbf{s}})\triangleq h_k(\mathbf{s}).
\end{equation} 
As a results, we can approximate \eqref{Received_SNR_Basic} as follows:
\begin{align}\label{Received_SNR}
	\gamma_k\approx\frac{\lvert\mathcal{A}_k\rvert}{\sigma_k^2}{\left\lvert\int_{{\mathcal{A}}}h_k({\mathbf{s}})j({\mathbf{s}})
		{\rm{d}}{\mathbf{s}}\right\rvert^2}.
\end{align} 
\subsection{Secure Transmission}\label{Section: System Model: Secure Transmission}
Bob makes an estimate $\hat{W}$ of $W$ based on the received output ${\mathbf{y}}_{\rm{b}}=[{{y}}_{\rm{b}}(1),\ldots,{{y}}_{\rm{b}}(N)]^{\mathsf{T}}$ from its channel, resulting in a block error rate $\epsilon_N=\Pr(W\ne \hat{W})$. The confidential message $W$ is also the input to Eve’s channel, and Eve has an average residual uncertainty $H(W|{\mathbf{y}}_{\rm{e}})$ after observing the output ${\mathbf{y}}_{\rm{e}}=[{{y}}_{\rm{e}}(1),\ldots,{{y}}_{\rm{e}}(N)]^{\mathsf{T}}$. Defining the transmission rate as $\mathcal{R}_N\triangleq H(W)/N$, and the fractional equivocation of Eve as $\Delta_N\triangleq H(W|{\mathbf{y}}_{\rm{e}})/H(W)$, the information-theoretic limits of secure transmission is described as follows \cite{leung1978gaussian}.
\subsubsection*{Secure Coding Theorem}
For any transmission rate ${\mathcal{R}}<{\mathcal{R}}_{j(\mathbf{s})}\triangleq\max\{\log_2(1+\gamma_{\rm{b}})-\log_2(1+\gamma_{\rm{e}}),0\}$, there exists an encoder-decoder pair such that as $N\rightarrow\infty$, the rate $\mathcal{R}_N\rightarrow\mathcal{R}$, the equivocation $\Delta_N\rightarrow1$, and the error probability $\epsilon_N\rightarrow0$.

We comment that $\Delta_N\rightarrow1$ is equivalent to $H(W|{\mathbf{y}}_{\rm{e}})\rightarrow H(W)$, or $I(W;{\mathbf{y}}_{\rm{e}})=H(W)-H(W|{\mathbf{y}}_{\rm{e}})\rightarrow0$, meaning that Eve is unable to extract any information about the confidential message $W$. The above statements suggest that the maximum secure transmission rate for a given source current distribution $j({\mathbf{s}})$ is ${\mathcal{R}}_{j(\mathbf{s})}$. Below this rate, Bob is able to recover the confidential message with arbitrary precision, while Eve cannot obtain any information from $W$.

Since the secrecy rate ${\mathcal{R}}_{j(\mathbf{s})}$ is a function of the current distribution $j({\mathbf{s}})$, we consider two key metrics to characterize the secrecy performance limits of the considered system.
\subsubsection{Maximum Secrecy Rate}
The MSR, subject to the power budget $P$, is defined as follows:
\begin{align}
{\mathcal{R}}_{\star}&\triangleq\max_{\int_{\mathcal{A}}\lvert j({\mathbf{s}})\rvert^2
{\rm{d}}{\mathbf{s}}\leq P}{\mathcal{R}}_{j(\mathbf{s})}\notag\\
&=\max_{\int_{\mathcal{A}}\lvert j({\mathbf{s}})\rvert^2
{\rm{d}}{\mathbf{s}}\leq P}\max\left\{\log_2\left(\frac{1+\gamma_{\rm{b}}}{1+\gamma_{\rm{e}}}\right),0\right\}.\label{Maximum_Rate_Definition}
\end{align}
Note that the secrecy rate in \eqref{Maximum_Rate_Definition} is an instantaneous metric evaluated for a given channel realization under the available CSI, rather than an ergodic average over fading.
\begin{remark}[Ergodic setting and CSI uncertainty]
If small-scale fading is present and ergodic averaging is of interest, a natural objective would be the ergodic secrecy rate $\mathbb{E}\left[\max\left\{\log_2\left(\frac{1+\gamma_{\rm{b}}}{1+\gamma_{\rm{e}}}\right),0\right\}\right]$. Moreover, if the CSI (especially Eve’s CSI) is imperfect, the design typically shifts to robust counterparts, e.g., maximizing a worst-case secrecy metric over an uncertainty set or adopting chance-constrained/stochastic robust designs. These extensions change the optimization objective/constraints and are beyond the scope of this paper.
\end{remark}

\subsubsection{Minimum Required Transmit Power}
Another metric of interest is the MRP to ensure a target secrecy rate ${\mathsf{R}}_0>0$, which is defined as follows:
\begin{align}
{\mathcal{P}}_{\star}&\triangleq\min_{{\mathcal{R}}_{j(\mathbf{s})}\geq {\mathsf{R}}_0}{\int_{\mathcal{A}}\lvert j({\mathbf{s}})\rvert^2
{\rm{d}}{\mathbf{s}}}\notag\\&=\min_{\log_2\left(\frac{1+\gamma_{\rm{b}}}{1+\gamma_{\rm{e}}}\right)\geq {\mathsf{R}}_0}{\int_{\mathcal{A}}\lvert j({\mathbf{s}})\rvert^2
{\rm{d}}{\mathbf{s}}}.\label{Minimum_Power_Definition}
\end{align}
Achieving the MSR or the MRP are fundamental objectives in secure transmission. The MSR represents the highest rate at which data can be securely transmitted, while the MRP indicates the minimum power needed to achieve a desired level of security. In the following sections, we derive closed-form expressions for ${\mathcal{R}}_{\star}$ and ${\mathcal{P}}_{\star}$, as defined in \eqref{Maximum_Rate_Definition} and \eqref{Minimum_Power_Definition}, respectively, and analyze these two key metrics.
\section{Analysis of the Maximum Secrecy Rate}\label{sec_MSR}
In this section, we analyze the MSR by deriving its close-form expression and the associate current distribution.
\subsection{Problem Reformulation}
Based on \eqref{Received_SNR} and \eqref{Maximum_Rate_Definition}, the problem of secrecy rate maximization can be formulated as follows:
\begin{equation}\label{CAP_MISOSE_Problem_1}
\max_{\int_{\mathcal{A}}{\left| j(\mathbf{s}) \right|^2}\mathrm{d}\mathbf{s}\le P} \frac{1+\frac{\lvert\mathcal{A}_{\mathrm{b}}\rvert}{\sigma _{\mathrm{b}}^{2}}\left| \int_{\mathcal{A}}{h_{\mathrm{b}}}(\mathbf{s})j(\mathbf{s})\mathrm{d}\mathbf{s} \right|^2}{1+\frac{\lvert\mathcal{A}_{\mathrm{e}}\rvert}{\sigma _{\mathrm{e}}^{2}}\left| \int_{\mathcal{A}}{h_{\mathrm{e}}}(\mathbf{s})j(\mathbf{s})\mathrm{d}\mathbf{s} \right|^2}.
\end{equation}
According to the monotonicity of function $f(x)=\frac{1+ax}{1+bx}$ for $x>0$ and $a>b>0$, it can be easily shown that the optimal $j(\mathbf{s})$ satisfies $\int_{\mathcal{A}}{\left| j(\mathbf{s}) \right|^2}\mathrm{d}\mathbf{s}= P$. By defining $\overline{\gamma}_{k}\triangleq\frac{P\lvert\mathcal{A}_k\rvert}{{\sigma}_k^2}
$ for $k\in\{\rm{b},\rm{e}\}$ and $u(\mathbf{s})\triangleq \frac{j(\mathbf{s})}{\sqrt{P}}$, we rewrite \eqref{CAP_MISOSE_Problem_1} as follows:
\begin{equation}\label{CAP_MISOSE_Problem_2}
\max_{\int_{\mathcal{A}}{\left| u(\mathbf{s}) \right|^2}\mathrm{d}\mathbf{s}=1} \frac{1+\overline{\gamma }_{\mathrm{b}}\left| \int_{\mathcal{A}}h_{\mathrm{b}}(\mathbf{s})u(\mathbf{s})\mathrm{d}\mathbf{s} \right|^2}{1+\overline{\gamma }_{\mathrm{e}}\left| \int_{\mathcal{A}}{h_{\mathrm{e}}}(\mathbf{s})u(\mathbf{s})\mathrm{d}\mathbf{s} \right|^2}.
\end{equation}
Furthermore, for $k\in\{\rm{b},\rm{e}\}$, it holds that
\begin{subequations}
	\begin{align}
		&1=\int_{\mathcal{A}}{\left\lvert u(\mathbf{s}) \right\rvert^2}\mathrm{d}\mathbf{s}\overset{\clubsuit}{=}\int_{\mathcal{A}}\int_{\mathcal{A}}u(\mathbf{s})\delta (\mathbf{s}-\mathbf{s}')u^*(\mathbf{s}')\mathrm{d}\mathbf{s}\mathrm{d}\mathbf{s}',\\
		&\left\lvert \int_{\mathcal{A}}{h_k(\mathbf{s})u(\mathbf{s})}\mathrm{d}\mathbf{s} \right\rvert^2=\int_{\mathcal{A}}\int_{\mathcal{A}}{{u(\mathbf{s})h_k(\mathbf{s})h_{k}^{*}(\mathbf{s}')u^*(\mathbf{s}')\mathrm{d}\mathbf{s}\mathrm{d}\mathbf{s}'}},
	\end{align}
\end{subequations}
where $\delta(\cdot)$ is the Dirac delta function, step $\clubsuit$ follows from the fact that $\int_{{\mathcal{A}}}\delta({\mathbf{x}}-{\mathbf{x}}_0)f(\mathbf{x}){\rm{d}}{\mathbf{x}}=f({\mathbf{x}}_0)$ with $f(\cdot)$ being an arbitrary function defined on ${\mathcal{A}}$. Taken together, we obtain
\begin{align}
1+\overline{\gamma }_k\left\lvert \int_{\mathcal{A}}{h_k(\mathbf{s})u(\mathbf{s})}\mathrm{d}\mathbf{s} \right\rvert^2
\!=\int_{\mathcal{A}}\int_{\mathcal{A}}{{u(\mathbf{s})A_k(\mathbf{s},\mathbf{s}^{\prime})}}u^*(\mathbf{s}')\mathrm{d}\mathbf{s}\mathrm{d}\mathbf{s}',
\end{align}
where 
\begin{equation}
	A_k(\mathbf{s},\mathbf{s}^{\prime})\triangleq  \delta (\mathbf{s}-\mathbf{s}^\prime)+\overline{\gamma }_{k}h_{k}(\mathbf{s})h_{k}^{*}(\mathbf{s}^\prime)
\end{equation}
Notably, we work on the Hilbert space $\mathcal{H}=L^2(\mathcal{A})$, where $L^2(\mathcal{A})$ is the Hilbert space of square-integrable functions on $\mathcal{A}$ defined as $L^2\left( \mathcal{A} \right) \triangleq \left\{ f:\mathcal{A} \rightarrow \mathbb{C} \left| \int_{\mathcal{A}}{\left| f\left( \mathbf{s} \right) \right|^2\mathrm{d}\mathbf{s}}<\infty \right.  \right\}$. Accordingly, all operators in this section are understood as bounded linear operators on $\mathcal{H}$. Further, it is noted that $A_k(\mathbf{s},\mathbf{s}^{\prime})$ is a rank-one perturbation of identity, which is compact (finite-rank) and self-adjoint. 

Consequently, the objective function in \eqref{CAP_MISOSE_Problem_2} equals the generalized Rayleigh quotient given as follows:
\begin{equation}\label{objective}
{\mathsf{RQ}}(u(\mathbf{s}))\triangleq\frac{\int_{\mathcal{A}}\int_{\mathcal{A}}{u(\mathbf{s})A_{\rm{b}}(\mathbf{s},\mathbf{s}^{\prime})}u^*(\mathbf{s}')\mathrm{d}\mathbf{s}\mathrm{d}\mathbf{s}'}
{\int_{\mathcal{A}}{\int_{\mathcal{A}}{u(\mathbf{s})A_{\rm{e}}(\mathbf{s},\mathbf{s}^{\prime})}}u^*(\mathbf{s}')\mathrm{d}\mathbf{s}\mathrm{d}\mathbf{s}'}.
\end{equation}
We note that the value of ${\mathsf{RQ}}(u(\mathbf{s}))$ is not affected by the norm of $u(\mathbf{s})$, i.e., $\int_{\mathcal{A}}{\left| u(\mathbf{s}) \right|^2\mathrm{d}\mathbf{s}}$. Therefore, the optimization problem in \eqref{CAP_MISOSE_Problem_1} is equivalent to the following one:
\begin{equation}\label{objective_trans_final}
\max_{u(\mathbf{s})}{\mathsf{RQ}}(u(\mathbf{s}))\Leftrightarrow\eqref{CAP_MISOSE_Problem_1}.
\end{equation}

To solve problem \eqref{objective_trans_final}, we define two functions as follows: 
\begin{align}
	Q\left( \mathbf{s},\mathbf{s}' \right)& \triangleq\delta (\mathbf{s}-\mathbf{s}')+\mu h_{\mathrm{e}}(\mathbf{s})h_{\mathrm{e}}^{*}(\mathbf{s}'),\label{function_Q}\\
	\hat{Q}\left( \mathbf{s},\mathbf{s}' \right)& \triangleq \delta (\mathbf{s}-\mathbf{s}')-\frac{\mu}{1+\mu g_{\mathrm{e}}}h_{\mathrm{e}}(\mathbf{s})h_{\mathrm{e}}^{*}(\mathbf{s}'),\label{Q_invert}
\end{align}
where $\mu =-\frac{1}{g_{\mathrm{e}}}+ \frac{1}{g_{\mathrm{e}}\sqrt{1+\overline{\gamma }_{\mathrm{e}}g_{\mathrm{e}}}}$, and $g_{\mathrm{e}}=\int_{\mathcal{A}}{\left| h_{\mathrm{e}}(\mathbf{s}) \right|^2\mathrm{d}\mathbf{s}}$ represents the channel gain for Eve. Then, the following lemmas can be found.
\vspace{-5pt}
\begin{lemma}\label{lemma_1}
Given $\gamma _{\mathrm{e}}>0$ and $g_{\mathrm{e}}>0$, $Q\left( \mathbf{s},\mathbf{s}' \right)$ and $\hat{Q}\left( \mathbf{s},\mathbf{s}' \right)$ are mutually invertible, which satisfy
\begin{equation}\label{Q_Inversion}
\begin{split}
\int_{\mathcal{A}}Q\left( \mathbf{s},\mathbf{s}_1 \right)\hat{Q}\left( \mathbf{s}_1,\mathbf{s}' \right){\rm{d}}\mathbf{s}_1
&=\int_{\mathcal{A}}\hat{Q}\left( \mathbf{s},\mathbf{s}_1 \right)Q\left( \mathbf{s}_1,\mathbf{s}' \right){\rm{d}}\mathbf{s}_1\\
&=\delta(\mathbf{s}-\mathbf{s}').
\end{split}
\end{equation}
\end{lemma}
\vspace{-5pt}
\begin{IEEEproof}
Please refer to Appendix \ref{Appendix:A} for more details.
\end{IEEEproof}
\vspace{-5pt}
\begin{lemma}\label{lemma_2}
Given $h_{\mathrm{e}}(\mathbf{s})$, it holds that
\begin{equation}\label{Indentity_Transform}
\begin{split}
\int_{\mathcal{A}}\!\int_{\mathcal{A}}\!Q\left( \mathbf{s}_1,\mathbf{s} \right)A_{\rm{e}}(\mathbf{s},\mathbf{s}')
Q\left( \mathbf{s}' ,\mathbf{s}_1'\right){\rm{d}}\mathbf{s}{\rm{d}}\mathbf{s}'=\delta(\mathbf{s}_1-\mathbf{s}_1').
\end{split}
\end{equation}
\end{lemma}
\begin{IEEEproof}
Please refer to Appendix \ref{Appendix:B} for more details.
\end{IEEEproof}
Motivated by the results in \textbf{Lemma \ref{lemma_2}}, we define the following function:
\begin{align}\label{Transform_Secrecy_Rate_Current}
{\nu }(\mathbf{s}_1)\triangleq\int_{{\mathcal{A}}}u({\mathbf{s}})\hat{Q}\left( \mathbf{s},\mathbf{s}_1 \right){\rm{d}}{\mathbf{s}},
\end{align}
which, together with \eqref{Q_Inversion}, yields
\begin{align}
\int_{\mathcal{A}}{\nu }(\mathbf{s}_1)Q\left( \mathbf{s}_1,\mathbf{s}' \right){\rm{d}}{\mathbf{s}}_1
&=\int_{{\mathcal{A}}}\int_{\mathcal{A}}u({\mathbf{s}})\hat{Q}\left( \mathbf{s},\mathbf{s}_1 \right)Q\left( \mathbf{s}_1,\mathbf{s}' \right){\rm{d}}{\mathbf{s}}\nonumber\\
&=\int_{\mathcal{A}}u({\mathbf{s}})\delta(\mathbf{s}-\mathbf{s}'){\rm{d}}{\mathbf{s}}=u(\mathbf{s}').
\end{align}
The above arguments imply that $u(\cdot)$ and ${\nu }(\cdot)$ can be mutually transformed. Therefore, we can transform the variable to be optimized in \eqref{objective_trans_final} from $u(\mathbf{s})$ to ${\nu }(\mathbf{s})$ by setting $u(\mathbf{s})=\int_{\mathcal{A}}{\nu }(\mathbf{s}_1)Q\left({\mathbf{s}}_1,\mathbf{s}\right){\rm{d}}{\mathbf{s}}_1$. As a result, the denominator of \eqref{objective} can be written as follows: 
\begin{equation}\label{Dominator_Transformed}
\begin{split}
&\int_{\mathcal{A}}{\int_{\mathcal{A}}{u(\mathbf{s})A_{\rm{e}}(\mathbf{s},\mathbf{s}^{\prime})}}u^*(\mathbf{s}')\mathrm{d}\mathbf{s}\mathrm{d}\mathbf{s}'\\
&=\int_{\mathcal{A}}\int_{\mathcal{A}}{\nu }(\mathbf{s}_1){\hat{E}}(\mathbf{s}_1,\mathbf{s}_1'){\nu }^{*}(\mathbf{s}_1')\mathrm{d}\mathbf{s}_1\mathrm{d}\mathbf{s}_1',
\end{split}
\end{equation}
where ${\hat{E}}(\mathbf{s}_1,\mathbf{s}_1')=\int_{\mathcal{A}}\int_{\mathcal{A}}Q\left({\mathbf{s}}_1,\mathbf{s}\right)A_{\rm{e}}(\mathbf{s},\mathbf{s}^{\prime}) 
Q^{*}\left({\mathbf{s}}_1',\mathbf{s}'\right)\mathrm{d}\mathbf{s}\mathrm{d}\mathbf{s}'$. According to the definition of $Q(\cdot,\cdot)$, we have $Q^{*}\left({\mathbf{s}}_1',\mathbf{s}'\right)=Q\left(\mathbf{s}',{\mathbf{s}}_1'\right)$, which, together with \textbf{Lemma \ref{lemma_1}}, yields ${\hat{E}}(\mathbf{s}_1,\mathbf{s}_1')=\delta(\mathbf{s}_1,\mathbf{s}_1')$. It follows that
\begin{align}
\int_{\mathcal{A}}\int_{\mathcal{A}}{\nu }(\mathbf{s}_1)\delta(\mathbf{s}_1,\mathbf{s}_1'){\nu }^{*}(\mathbf{s}_1')\mathrm{d}\mathbf{s}_1\mathrm{d}\mathbf{s}_1'
=\int_{\mathcal{A}}{\left| \nu (\mathbf{s}_1) \right|^2\mathrm{d}\mathbf{s}_1}.
\end{align}
Taken together, we transform problem \eqref{objective_trans_final} into the following equivalent form:
\begin{align}\label{CAP_MISOSE_Problem_4}
\max_{\nu (\mathbf{s}_1)}\frac{\int_{\mathcal{A}}{\int_{\mathcal{A}}}\nu (\mathbf{s}_1)B(\mathbf{s}_1,\mathbf{s}_1^{\prime})\nu ^*(\mathbf{s}_1^{\prime})\mathrm{d}\mathbf{s}_1\mathrm{d}\mathbf{s}_1^{\prime}}{\int_{\mathcal{A}}{\left| \nu (\mathbf{s}_1) \right|^2}\mathrm{d}\mathbf{s}_1},
\end{align}
where 
\begin{equation}\label{Semi_Def_Function_Original}
B(\mathbf{s}_1,\mathbf{s}_1^{\prime})=\int_{\mathcal{A}}\int_{\mathcal{A}}Q\left( \mathbf{s}_1,\mathbf{s} \right)A_{\rm{b}}(\mathbf{s},\mathbf{s}')
Q\left( \mathbf{s}' ,\mathbf{s}_1'\right){\rm{d}}\mathbf{s}{\rm{d}}\mathbf{s}'.
\end{equation}
Note that the objective function of problem \eqref{CAP_MISOSE_Problem_4} is not influenced by the norm of $\nu(\mathbf{s}_1)$, i.e., $\int_{\mathcal{A}}{\left| \nu(\mathbf{s}_1) \right|^2\mathrm{d}\mathbf{s}_1}$. Thus, problem \eqref{CAP_MISOSE_Problem_4} can be simplified by removing the denominator term therein. The final results are given as follows. 
\vspace{-5pt}
\begin{theorem}\label{Theorem_CAP_MISOSE_Problem}
The secrecy rate maximization problem defined in \eqref{CAP_MISOSE_Problem_1} is equivalent to the following:
\begin{equation}\label{CAP_MISOSE_Problem_5}
\nu^{\star}(\mathbf{s}_1)=\argmax_{\nu (\mathbf{s}_1)} \int_{\mathcal{A}}{\int_{\mathcal{A}}}{\nu } (\mathbf{s}_1)B(\mathbf{s}_1,\mathbf{s}_1^{\prime}){\nu } ^*(\mathbf{s}_1^{\prime})\mathrm{d}\mathbf{s}_1\mathrm{d}\mathbf{s}_1^{\prime}.
\end{equation}
After obtaining $\nu^{\star}(\mathbf{s}_1)$, the source current distribution that maximizes the secrecy rate can be calculated as follows:
\begin{align}\label{optimal_u_rate_final_current}
j_{\mathsf{msr}}\left( \mathbf{s} \right)=\frac{\sqrt{P}\int_{\mathcal{A}}{\nu }^{\star}(\mathbf{s}_1)Q\left({\mathbf{s}}_1,\mathbf{s}\right){\rm{d}}{\mathbf{s}}_1}
{\sqrt{\int_{{\mathcal{A}}}\lvert\int_{\mathcal{A}}{\nu }^{\star}(\mathbf{s}_1)Q\left({\mathbf{s}}_1,\mathbf{s}\right){\rm{d}}{\mathbf{s}}_1\rvert^2{\rm{d}}{\mathbf{s}}}}.
\end{align}
\end{theorem}
\vspace{-5pt}
\begin{IEEEproof}
The results follow directly from using \eqref{Transform_Secrecy_Rate_Current}.
\end{IEEEproof}
\subsection{Maximum Secrecy Rate \& Optimal Current Distribution}
In this subsection, we aim to solve problem \eqref{CAP_MISOSE_Problem_5} and derive closed-form expressions for the optimal current distribution and the achieved MSR. 
\subsubsection{Maximum Secrecy Rate}
The results in \textbf{Theorem \ref{Theorem_CAP_MISOSE_Problem}} suggest that the optimal solution to problem \eqref{CAP_MISOSE_Problem_5}, i.e., the rate-optimal source current distribution, corresponds to the principal eigenfunction of the operator $B(\mathbf{s}_1,\mathbf{s}_1^{\prime})$. Furthermore, the optimal objective value of problem \eqref{CAP_MISOSE_Problem_5} is given by the principal eigenvalue of $B(\mathbf{s}_1,\mathbf{s}_1^{\prime})$. Therefore, we have
\begin{equation}\label{maximum_secrecy_rate_definition_final}
	{\mathcal{R}}_{\star}=\max_{\int_{\mathcal{A}}\lvert j({\mathbf{s}})\rvert^2
{\rm{d}}{\mathbf{s}}\leq P}{\mathcal{R}}_{j(\mathbf{s})}=\log_2(\lambda_{B}^{\max}),
\end{equation}
where $\lambda _{B}^{\max}$ denotes the principal eigenvalue of $B(\mathbf{s}_1,\mathbf{s}_1^{\prime})$\footnote{The operator $B(\cdot,\cdot)$ defined in \eqref{Semi_Def_Function_Original} is self-joint on $\mathcal{H}$ and has the structure “identity plus a finite-rank (hence compact) perturbation” inherited from $A_\mathrm{b}(\cdot,\cdot)$ and $Q(\cdot,\cdot)$. Consequently, its spectrum consists of the essential part at $1$ and a finite number of isolated real eigenvalues with finite multiplicity, which justifies the existence of the largest eigenvalue $\lambda^{\mathrm{max}}_B$.}. 

To calculate each eigenvalue $\lambda_B$ of $B(\mathbf{s}_1,\mathbf{s}_1^{\prime})$, we need to solve the following characteristic equation:
\begin{equation}\label{character}
	\det(B(\mathbf{s}_1,\mathbf{s}_1^{\prime})-\lambda_B\delta(\mathbf{s}_1-\mathbf{s}_1^{\prime}))=0,
\end{equation}
where $\det(\cdot)$ here is utilized to calculate the Fredholm determinant of the operator $B(\mathbf{s}_1,\mathbf{s}_1^{\prime})-\lambda_B\delta(\mathbf{s}_1-\mathbf{s}_1^{\prime})\triangleq{\hat{B}}(\mathbf{s}_1,\mathbf{s}_1^{\prime})$, i.e., the product of all its eigenvalues \cite{gohberg2012traces}. Since directly solving equation \eqref{character} is challenging, we define
\begin{align}\label{C_definition}
	C(\mathbf{s}_2,\mathbf{s}_2^{\prime})\triangleq&\int_{\mathcal{A}}{\int_{\mathcal{A}}}\hat{Q}(\mathbf{s}_2,\mathbf{s}_1){\hat{B}}(\mathbf{s}_1,\mathbf{s}_1^{\prime})
	\hat{Q}(\mathbf{s}^\prime_1,\mathbf{s}^\prime_2)\mathrm{d}\mathbf{s}_1\mathrm{d}\mathbf{s}_{1}^{\prime}.
\end{align}
Then, the following lemma can be concluded.
\vspace{-5pt}
\begin{lemma}\label{Lemma_Fredholm}
Given $h_{\mathrm{b}}(\mathbf{s})$ and $h_{\mathrm{e}}(\mathbf{s})$, it holds that 
\begin{equation}\label{C_derivation}
\begin{split}
	C(\mathbf{s}_2,\mathbf{s}_2^{\prime})&=\overline{\gamma }_{\mathrm{b}}h_{\mathrm{b}}(\mathbf{s}_2)h_{\mathrm{b}}^{*}(\mathbf{s}_{2}^{\prime})-\lambda _B \overline{\gamma }_{\mathrm{e}} h_{\mathrm{e}}(\mathbf{s}_2)h_{\mathrm{e}}^{*}(\mathbf{s}_{2}^{\prime})\\
	&+(1-\lambda _B)\delta (\mathbf{s}_2-\mathbf{s}_{2}^{\prime}).
\end{split}
\end{equation}
\end{lemma}
\vspace{-5pt}
\begin{IEEEproof}
Please refer to Appendix \ref{Appendix:C} for more details.
\end{IEEEproof}
According to \textbf{Lemma \ref{lemma_2}}, $\hat{Q}(\cdot,\cdot)$ is an invertible operator. Consequently, we have
\begin{equation}
	\det({\hat{B}}(\mathbf{s}_1,\mathbf{s}_1^{\prime}))=0\Leftrightarrow\det(C(\mathbf{s}_2,\mathbf{s}_2^{\prime}))=0,
\end{equation}
which means that we can equivalently transform equation \eqref{character} to the following equation:
\begin{equation}\label{equation_trans}
\det(C(\mathbf{s}_2,\mathbf{s}_2^{\prime}))	=0.
\end{equation}

Note that the operator $C(\mathbf{s}_2,\mathbf{s}_2^{\prime})$ is Hermitian, i.e., $C(\mathbf{s}_2,\mathbf{s}_2^{\prime})=C^*(\mathbf{s}_2^{\prime},\mathbf{s}_2)$, and thus its determinant is equal to the product of the eigenvalues, i.e., $\det(C(\mathbf{s}_2,\mathbf{s}_2^{\prime}))=\prod_{i=1}^{\infty}{\lambda _{C,i}}$ with $\lambda _{C,i}$ being the $i$th eigenvalue of $C(\mathbf{s}_2,\mathbf{s}_2^{\prime})$. Consequently, we can rewrite \eqref{equation_trans} as follows:
\begin{equation}\label{equation_trans2}
	\det(C(\mathbf{s}_2,\mathbf{s}_2^{\prime}))	=\prod\nolimits_{i=1}^{\infty}{\lambda _{C,i}}=0.
\end{equation}
By observing the mathematical structure of $C(\mathbf{s}_2,\mathbf{s}_2^{\prime})$ given in \eqref{C_derivation}, we conclude the following lemma.
\vspace{-5pt}
\begin{lemma}\label{lem_eigenvalue}
Given $h_{\mathrm{b}}(\mathbf{s})$ and $h_{\mathrm{e}}(\mathbf{s})$, the eigenvalues of $C(\mathbf{s}_2,\mathbf{s}_2^{\prime})$	are given as follows:
\begin{subequations}\label{eigenvalue_C}
	\begin{align}
		&\lambda_{C,1}=\xi_1-\lambda_B+1,\ \lambda_{C,2}=\xi_2-\lambda_B+1,\\
		&\lambda_{C,3}=\ldots=\lambda_{C,\infty}=-\lambda_B+1.
	\end{align}
\end{subequations}
Here,
\begin{subequations}
\begin{align}
	\xi_1&=\frac{\Delta+
		\sqrt{\Delta^2+4\lambda _B \overline{\gamma}_{\mathrm{b}}\overline{\gamma}_{\mathrm{e}}g_{\mathrm{b}}g_{\mathrm{e}}(1-\overline{\rho})}}{2},\label{eigenvalue_C_solution1}\\
	\xi_2&=\frac{\Delta-
		\sqrt{\Delta^2+4\lambda _B \overline{\gamma}_{\mathrm{b}}\overline{\gamma}_{\mathrm{e}}g_{\mathrm{b}}g_{\mathrm{e}}(1-\overline{\rho})}}{2},\label{eigenvalue_C_solution2}
\end{align}
\end{subequations}
 where $\Delta=\overline{\gamma}_{\mathrm{b}}g_{\mathrm{b}}-\lambda _B \overline{\gamma}_{\mathrm{e}}g_{\mathrm{e}}$, $g_{\mathrm{b}}=\int_{\mathcal{A}}{\left| h_{\mathrm{b}}(\mathbf{s}) \right|^2\mathrm{d}\mathbf{s}}$ represents the channel gain for Bob, $\overline{\rho }=\frac{\left| \rho \right|^2}{g_{\mathrm{b}}g_{\mathrm{e}}}\in[ 0,1 ) $\footnote{We assume that the spatial responses of Bob and Eve are not parallel, i.e., $\overline{\rho }\ne1$. This condition arises when Bob and Eve are located at different positions, which represents the most general and practical scenario.} denotes the normalized correlation factor between Bob and Eve, and $\rho \triangleq\int_{\mathcal{A}}{h_{\mathrm{b}}(\mathbf{s})h_{\mathrm{e}}^{*}(\mathbf{s})\mathrm{d}\mathbf{s}}$ is the channel correlation (inner product).
\end{lemma}
\vspace{-3pt}
\begin{IEEEproof}
Please refer to Appendix \ref{Appendix:D} for more details.
\end{IEEEproof}
Substituting \eqref{eigenvalue_C} into \eqref{equation_trans2} gives
\begin{equation}\label{equation_trans3}
	(\xi_1-\lambda_B+1)(\xi_2-\lambda_B+1)\prod\nolimits_{i=3}^{\infty}(-\lambda_B+1)=0.
\end{equation}
By solving the above equation with respect to $\lambda_B$, we can obtain all the eigenvalues of $B(\mathbf{s}_1,\mathbf{s}_1^{\prime})$. The corresponding results are summarized as follows.
\vspace{-5pt}
\begin{theorem}\label{lem_principal}
Given $h_{\mathrm{b}}(\mathbf{s})$ and $h_{\mathrm{e}}(\mathbf{s})$, the principal eigenvalue of operator $B(\mathbf{s}_1,\mathbf{s}_1^{\prime})$ can be expressed as follows:
\begin{equation}\label{max_secrecy_rate_maximum_eigenvalue}
\lambda _{B}^{\max}=1+\frac{\overline{\gamma }_{\mathrm{b}}g_{\mathrm{b}}\left( 1+\overline{\gamma }_{\mathrm{e}}g_{\mathrm{e}}\left( 1-\overline{\rho } \right) \right)}{1+\overline{\gamma }_{\mathrm{e}}g_{\mathrm{e}}}.
\end{equation}
\end{theorem}
\vspace{-5pt}
\begin{IEEEproof}
Please refer to Appendix \ref{Appendix:E} for more details.
\end{IEEEproof}
Consequently, we derive the closed-form expression for the MSR in the following theorem.
\vspace{-5pt}
\begin{theorem}
Given $h_{\mathrm{b}}(\mathbf{s})$ and $h_{\mathrm{e}}(\mathbf{s})$, the MSR is given by
\begin{equation}\label{max_secrecy_rate}
{\mathcal{R}}_{\star}=\log_2\left(1+\frac{\overline{\gamma }_{\mathrm{b}}g_{\mathrm{b}}\left( 1+\overline{\gamma }_{\mathrm{e}}g_{\mathrm{e}}\left( 1-\overline{\rho } \right) \right)}{1+\overline{\gamma }_{\mathrm{e}}g_{\mathrm{e}}}\right)
\triangleq{R}(g_{\mathrm{b}},g_{\mathrm{e}},\overline{\rho }).
\end{equation}
\end{theorem}
\vspace{-5pt}
\begin{IEEEproof}
The results follow by inserting \eqref{max_secrecy_rate_maximum_eigenvalue} into \eqref{maximum_secrecy_rate_definition_final}.
\end{IEEEproof}
\vspace{-5pt}
\begin{remark}\label{rem_gen}
The results in \eqref{max_secrecy_rate} suggest that the MSR is determined by the channel gain of each user and their channel correlation factor. This expression is applicable to any aperture, regardless of its location, shape, and size. Further, the above derivations are not specific to any particular channel and can be directly extended to various channel types.
\end{remark}

\subsubsection{Optimal Current Distribution}
Having calculated the MSR ${\mathcal{R}}_{\star}$, we turn to the source current that achieves it. 
\vspace{-5pt}
\begin{lemma}\label{lem_optimal_u}
The optimal solution to the simplified secrecy rate maximization problem defined in \eqref{CAP_MISOSE_Problem_5} is given by
\begin{align}\label{optimal_u_rate_step0}
\nu^{\star}(\mathbf{s}_1)=\int_{{\mathcal{A}}}u^{\star}({\mathbf{s}})\hat{Q}\left( \mathbf{s},\mathbf{s}_1 \right){\rm{d}}{\mathbf{s}},
\end{align}
where $u^\star\left( \mathbf{s} \right)$ is the principal eigenfunction of $C(\mathbf{s},\mathbf{s}^{\prime})$.
\end{lemma}
\vspace{-5pt}
\begin{IEEEproof}
Please refer to Appendix \ref{Appendix:F} for more details.
\end{IEEEproof}
Following the derivation steps outlined in Appendix \ref{Appendix:D}, the principal eigenfunction of $C(\mathbf{s},\mathbf{s}^{\prime})$ is given by
\begin{equation}\label{optimal_u_rate_step1}
	u^{\star}\left( \mathbf{s} \right) ={h^*_{\mathrm{b}}(\mathbf{s})+\frac{\xi_1-\overline{\gamma}_{\mathrm{b}}g_{\mathrm{b}}}{\overline{\gamma}_{\mathrm{b}}\rho}h^*_{\mathrm{e}}(\mathbf{s})},
\end{equation}
which is obtained by substituting the principal eigenvalue $\xi=\xi_1$ into \eqref{b_a}. By submitting \eqref{optimal_u_rate_step0} and \eqref{optimal_u_rate_step1} into \eqref{optimal_u_rate_final_current} and using the results in \textbf{Lemma \ref{lemma_2}} for simplifications, the optimal current distribution that maximizes the secrecy rate is given in the following theorem.
\vspace{-5pt}
\begin{theorem}
The optimal source current distribution that maximizes the secrecy transmission rate is given by
	\begin{equation}\label{optimal_j}
		j_{\mathsf{msr}}\left( \mathbf{s} \right)=\sqrt{P}\frac{h^*_{\mathrm{b}}(\mathbf{s})+\frac{\xi_1-\overline{\gamma}_{\mathrm{b}}g_{\mathrm{b}}}{\overline{\gamma}_{\mathrm{b}}\rho}h^*_{\mathrm{e}}(\mathbf{s})}
{\sqrt{\int_{\mathcal{A}}\left\lvert h^*_{\mathrm{b}}(\mathbf{s})+\frac{\xi_1-\overline{\gamma}_{\mathrm{b}}g_{\mathrm{b}}}{\overline{\gamma}_{\mathrm{b}}\rho}h^*_{\mathrm{e}}(\mathbf{s})\right\rvert^2{\rm{d}}{\mathbf{s}}}}.
	\end{equation}
\end{theorem}
\vspace{-5pt}
\subsubsection{Asymptotic Analysis}\label{Section: Asymptotic Analysis}
To gain further insights into the system, we conduct an asymptotic analysis of the MSR in both the low-SNR and high-SNR regimes.

In the low-SNR regime, i.e., $P\rightarrow0$, we have $\overline{\gamma }_{\mathrm{b}}\rightarrow0$ and $\overline{\gamma }_{\mathrm{e}}\rightarrow0$. According to the results in {\textbf{Theorem \ref{lem_principal}}}, we obtain
\begin{align}\label{Asymptotic_Analysis_Important_Property}
\xi_1-\lambda _{B}^{\max}+1=0\Rightarrow\xi_1=
\frac{\overline{\gamma }_{\mathrm{b}}g_{\mathrm{b}}\left( 1+\overline{\gamma }_{\mathrm{e}}g_{\mathrm{e}}\left( 1-\overline{\rho } \right) \right)}{1+\overline{\gamma }_{\mathrm{e}}g_{\mathrm{e}}}.
\end{align}
It follows that
\begin{align}
\lim_{P\rightarrow0}\frac{\xi_1}{\overline{\gamma}_{\mathrm{b}}\rho}=
\lim_{P\rightarrow0}\frac{g_{\mathrm{b}}\left( 1+\overline{\gamma }_{\mathrm{e}}g_{\mathrm{e}}\left( 1-\overline{\rho } \right) \right)}{\rho(1+\overline{\gamma }_{\mathrm{e}}g_{\mathrm{e}})}=\frac{g_{\mathrm{b}}}{\rho},
\end{align}
which, together with the fact that $\lim_{P\rightarrow0}\frac{\overline{\gamma}_{\mathrm{b}}g_{\mathrm{b}}}{\overline{\gamma}_{\mathrm{b}}\rho}=\frac{g_{\mathrm{b}}}{\rho}$, yields
\begin{align}
\lim_{P\rightarrow0}\frac{\xi_1-\overline{\gamma}_{\mathrm{b}}g_{\mathrm{b}}}{\overline{\gamma}_{\mathrm{b}}\rho}=\frac{g_{\mathrm{b}}}{\rho}-\frac{g_{\mathrm{b}}}{\rho}=0.
\end{align}
Therefore, in the low-SNR regime, the optimal source current distribution degenerates into the following form:
\begin{align}\label{Optimal_Current_Rate_Low_SNR}
\lim_{P\rightarrow0}j_{\mathsf{msr}}\left( \mathbf{s} \right)\simeq\sqrt{P}\frac{h^*_{\mathrm{b}}(\mathbf{s})}
{\sqrt{\int_{\mathcal{A}}\left\lvert h^*_{\mathrm{b}}(\mathbf{s})\right\rvert^2{\rm{d}}{\mathbf{s}}}},
\end{align}
which is parallel to Bob's spatial response, i.e., $h^*_{\mathrm{b}}(\mathbf{s})$.
\vspace{-3pt}
\begin{remark}
The results in \eqref{Optimal_Current_Rate_Low_SNR} suggest that in the low-SNR regime, the rate-optimal source current distribution simplifies to MRT beamforming, which aims to maximize the legitimate user's signal power.
\end{remark}
\vspace{-3pt}
Based on \eqref{Optimal_Current_Rate_Low_SNR}, the low-SNR MSR is given by
\begin{align}\label{Optimal_Rate_Low_SNR}
\lim_{P\rightarrow0}{\mathcal{R}}_{\star}\simeq\log _2\left( \frac{1+\overline{\gamma }_{\mathrm{b}}g_{\mathrm{b}}}{1+\overline{\gamma }_{\mathrm{e}}g_{\mathrm{e}}\overline{\rho }} \right)=\mathcal{R} _{\mathsf{mrt}},
\end{align}
where $\mathcal{R} _{\mathsf{mrt}}$ represents the secrecy rate achieved by MRT beamforming, i.e., $j_{\mathsf{mrt}}\left( \mathbf{s} \right) =\sqrt{P}\frac{h^*_{\mathrm{b}}\left( \mathbf{s} \right)}{\sqrt{\int_{\mathcal{A}}{\left| h_{\mathrm{b}}\left( \mathbf{s} \right) \right|^2\mathrm{d}\mathbf{s}}}}$.

In the high-SNR regime, i.e., $P\rightarrow\infty$, we have $\overline{\gamma }_{\mathrm{b}}\rightarrow\infty$ and $\overline{\gamma }_{\mathrm{e}}\rightarrow\infty$, which, together with \eqref{Asymptotic_Analysis_Important_Property}, leads to
\begin{align}
\lim_{P\rightarrow\infty}\frac{\xi_1}{\overline{\gamma}_{\mathrm{b}}\rho}=
\lim_{P\rightarrow\infty}\frac{g_{\mathrm{b}}( 1\!+\!\overline{\gamma }_{\mathrm{e}}g_{\mathrm{e}}( 1\!-\!\overline{\rho }))}{\rho(1+\overline{\gamma }_{\mathrm{e}}g_{\mathrm{e}})}=\frac{g_{\mathrm{b}}}{\rho}( 1-\overline{\rho } ).
\end{align}
Recalling that $\lim_{P\rightarrow\infty}\frac{\overline{\gamma}_{\mathrm{b}}g_{\mathrm{b}}}{\overline{\gamma}_{\mathrm{b}}\rho}=\frac{g_{\mathrm{b}}}{\rho}$ yields
\begin{align}
\lim_{P\rightarrow\infty}\frac{\xi_1-\overline{\gamma}_{\mathrm{b}}g_{\mathrm{b}}}{\overline{\gamma}_{\mathrm{b}}\rho}=\frac{g_{\mathrm{b}}}{\rho}( 1-\overline{\rho } )-\frac{g_{\mathrm{b}}}{\rho}= -\frac{g_{\mathrm{b}}}{\rho}\overline{\rho }= -\frac{\rho^*}{g_{\mathrm{e}}}.
\end{align}
Therefore, in the high-SNR regime, the optimal source current distribution degenerates into the following form: \begin{align}\label{Optimal_Current_Rate_High_SNR}
\lim_{P\rightarrow\infty}j_{\mathsf{msr}}\left( \mathbf{s} \right)\simeq\sqrt{P}\frac{h^*_{\mathrm{b}}(\mathbf{s})-\frac{\rho^*}{g_{\mathrm{e}}}h^*_{\mathrm{e}}(\mathbf{s})}
{\sqrt{\int_{\mathcal{A}}\left\lvert h^*_{\mathrm{b}}(\mathbf{s})-\frac{\rho^*}{g_{\mathrm{e}}}h^*_{\mathrm{e}}(\mathbf{s})\right\rvert^2{\rm{d}}{\mathbf{s}}}}.
\end{align}
We note that 
\begin{subequations}
\begin{align}
&\int_{\mathcal{A}}\left(h^*_{\mathrm{b}}(\mathbf{s})-\frac{\rho^*}{g_{\mathrm{e}}}h^*_{\mathrm{e}}(\mathbf{s})\right)h_{\mathrm{e}}(\mathbf{s}){\rm{d}}{\mathbf{s}}
=\rho^{*}-\rho^{*}=0,\label{ZF_Rate_Condition1}\\
&\int_{\mathcal{A}}\left(h^*_{\mathrm{b}}(\mathbf{s})-\frac{\rho^*}{g_{\mathrm{e}}}h^*_{\mathrm{e}}(\mathbf{s})\right)h_{\mathrm{b}}(\mathbf{s}){\rm{d}}{\mathbf{s}}
=g_{\mathrm{b}}-\frac{\lvert \rho\rvert^2}{g_{\mathrm{e}}}=g_{\mathrm{b}}(1-\overline{\rho }),\label{ZF_Rate_Condition2}
\end{align}
\end{subequations}
which suggests that $h^*_{\mathrm{b}}(\mathbf{s})-\frac{\rho^*}{g_{\mathrm{e}}}h^*_{\mathrm{e}}(\mathbf{s})$ is orthogonal to $h_{\mathrm{e}}(\mathbf{s})$.

\begin{remark}
The above arguments imply that in the high-SNR regime, the rate-optimal source current distribution simplifies to ZF beamforming, which aims to minimize the information leakage to the eavesdropper.
\end{remark}

Based on \eqref{Optimal_Current_Rate_High_SNR}, the high-SNR MSR is given by
\begin{align}\label{Optimal_Rate_High_SNR}
\lim_{P\rightarrow\infty}{\mathcal{R}}_{\star}\simeq\log_2\left( \overline{\gamma }_{\mathrm{b}}g_{\mathrm{b}}\left( 1-\overline{\rho } \right) \right)=\mathcal{R} _{\mathsf{zf}},
\end{align}
where $\mathcal{R} _{\mathsf{zf}}$ represents the secrecy rate achieved by ZF beamforming, i.e., $j_{\mathsf{zf}}\left( \mathbf{s} \right) =\sqrt{P}\frac{h^*_{\mathrm{b}}(\mathbf{s})-\frac{\rho^*}{g_{\mathrm{e}}}h^*_{\mathrm{e}}(\mathbf{s})}
{\sqrt{\int_{\mathcal{A}}\left\lvert h^*_{\mathrm{b}}(\mathbf{s})-\frac{\rho^*}{g_{\mathrm{e}}}h^*_{\mathrm{e}}(\mathbf{s})\right\rvert^2{\rm{d}}{\mathbf{s}}}}$. Notably, when $\rho\approx 0$, i.e., when Bob’s and Eve’s channels are nearly orthogonal, the ZF case will reduce to MRT.  

{These findings reveal that MRT and ZF beamforming represent two extremes of optimal secure beamforming strategies. In other words, the optimal current distribution achieves a balance between enhancing the intended signal and minimizing information leakage.  At low SNR, noise dominates and the impact of leakage is limited, so maximizing Bob’s received power, i.e., MRT beamforming, is optimal. Conversely, at high SNR, where noise becomes negligible and information leakage becomes the limiting factor, the optimal strategy shifts toward minimizing Eve's received power, i.e., ZF beamforming.}

\subsection{Typical Cases}\label{discussion_MSR}
As stated in \textbf{Remark~\ref{rem_gen}}, the derived expression for the MSR is applicable to arbitrary apertures and channel types. In this subsection, we specialize the aperture $\mathcal{A}$ o specific cases, such as planar CAPAs and planar SPDAs. 

To facilitate theoretical investigations into fundamental performance limits and asymptotic behaviors, we consider LoS channels. In particular, the channel response between point $\mathbf{s}\in\mathcal{A}$ and user $k\in\{\rm{b},\rm{e}\}$ is modeled as follows \cite{zhao2024continuous}:
\begin{equation}\label{channel_model}
h_k\left( \mathbf{s} \right) =\frac{\mathrm{j}k_0\eta \mathrm{e}^{-\mathrm{j}k_0\left\| \mathbf{r}_k-\mathbf{s} \right\|}}{\sqrt{4\pi} \left\| \mathbf{r}_k-\mathbf{s} \right\|}\sqrt{\frac{\left| \mathbf{e}^{\mathsf{T}}(\mathbf{s}-\mathbf{r}_k) \right|}{\left\| \mathbf{r}_k-\mathbf{s} \right\|}},	
\end{equation}
Here, the term $\sqrt{\frac{\left| \mathbf{e}^{\mathsf{T}}(\mathbf{s}-\mathbf{r}_k) \right|}{\left\| \mathbf{r}_k-\mathbf{s} \right\|}}$ models the effect of the project aperture of the BS array,as indicated by the projection of the array's normal vector ${\mathbf{e}}\in{\mathbb{R}}^{3\times1}$ onto the wave propagation direction at point $\mathbf{s}$. Additionally, $\frac{\mathrm{j}k_0\eta \mathrm{e}^{-\mathrm{j}k_0\left\| \mathbf{r}_k-\mathbf{s} \right\|}}{4\pi \left\| \mathbf{r}_k-\mathbf{s} \right\|}$ represents the impact of free-space EM propagation \cite{ouyang2024impact}, where $\eta=120 \pi~\Omega$ denotes the impedance of free space, and $k_0=\frac{2\pi}{\lambda}$ with $\lambda$ being the wavelength represents the wavenumber \footnote{It is important to emphasize that although we adopt the LoS channel model as an illustrative example, the derived results in the previous section are not constrained to any specific channel type and remain applicable in a wide range of channel environments.}.

\begin{figure}[!t]
	\centering
	\includegraphics[height=0.23\textwidth]{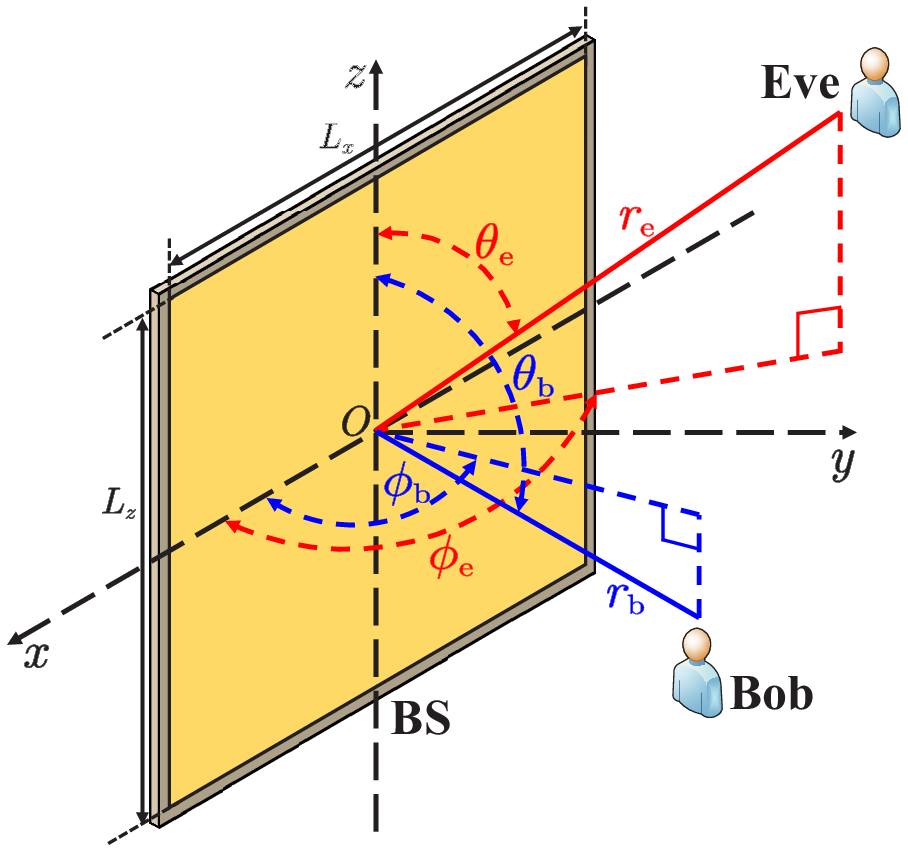}
	\caption{Illustration for a planar CAPA.}
	\vspace{-6pt}
	\label{planar_capa}
\end{figure}

\subsubsection{Planar CAPA}
We consider that the BS employs a planar CAPA situated on the $x$-$z$ plane and centered at the origin, as depicted in Fig. \ref{planar_capa}. The edges of $\mathcal{A}$ are parallel to the coordinate axes, with physical dimensions $L_x$ and $L_z$ along the $x$- and $z$-axes, respectively. It follows that
\begin{equation}
	\mathcal{A} \!=\!\{[x,0,z]^{\mathsf{T}}|x\!\in \![-{L_x}/{2},{L_x}/{2}],z\!\in \![-{L_z}/{2},{L_z}/{2}]\},
\end{equation} 
with the size $\lvert \mathcal{A} \rvert=L_xL_z$. For each user $k \in \{\rm{b}, \rm{e}\}$, let $r_k$ denote the distance from the center of $\mathcal{A}$ to the center of $\mathcal{A}_\mathsf{k}$, and let $\phi_k \in [0, \pi]$ and $\theta_k \in [0, \pi]$ represent the corresponding azimuth and elevation angles, respectively. Accordingly, $\mathcal{A}_\mathsf{k}$ is centered at ${\mathbf{r}}_{k}=[r_k\Phi_k, r_k\Psi_k, r_k\Theta_k]^{\mathsf{T}}$, where $\Phi_k\triangleq\cos{\phi_k}\sin{\theta_k}$, $\Psi_k\triangleq\sin{\phi_k}\sin{\theta_k}$, and $\Theta_k\triangleq\cos{\theta_k}$. Consequently, for $\mathbf{s}=[x,0,z]^{\mathsf{T}}\in\mathcal{A}$, by inserting $\mathbf{e}=[0,1,0]^\mathsf{T}$ and $\left\| \mathbf{r}_k-\mathbf{s} \right\| =(x^2+z^2-2r_k\left( \Phi _kx+\Theta _kz \right) +r_{k}^{2})^{\frac{1}{2}}$ into \eqref{channel_model}, we derive the channel response for user $k$ as follows:
\begin{equation}\label{case_stduy_G}
	\begin{split}
		h_k( \mathbf{s} ) &=\frac{\mathrm{j}k_0\eta \sqrt{r_k\Psi _k}\mathrm{e}^{-\mathrm{j}k_0(x^2+z^2-2r_k\left( \Phi _kx+\Theta_kz \right) +r_{k}^{2})^{\frac{1}{2}}}}{\sqrt{4\pi}(x^2+z^2-2r_k\left( \Phi _kx+\Theta _kz \right) +r_{k}^{2})^{\frac{3}{4}}}\\
		&\triangleq h_k(x,z).
	\end{split}		
\end{equation}
To characterize the MSR, we firstly derive the channel gain for each user and the channel correlation factor as follows.
\vspace{-5pt}
\begin{lemma}\label{lem_planar_capa}
With the planar CAPA, the channel gain for user $k\in\{{\mathrm{b},\mathrm{e}}\}$ can be expressed as follows:
\begin{equation}\label{g_planar_capa}
	g_k^{\mathrm{c}}=\frac{k_{0}^{2}\eta ^2}{4\pi}\sum_{x\in \mathcal{X} _k}{\sum_{z\in \mathcal{Z} _k}{\arctan}}\bigg( \frac{xz}{\Psi _k\sqrt{\Psi _{k}^{2}+x^2+z^2}} \bigg),
\end{equation}
where ${\mathcal{X}}_k\triangleq\{\frac{L_x}{2r_k}\pm \Phi_k\}$, and ${\mathcal{Z}}_k\triangleq\{\frac{L_z}{2r_k}\pm \Theta_k\}$. Additionally, the channel correlation factor can be approximated as follows:
\begin{align}\label{rho_planar_capa}
	\begin{split}
		\rho _{\mathrm{c}} =&\frac{\pi^2 \lvert \mathcal{A} \rvert}{4T^2}\sum\nolimits_{t=1}^T{\sum\nolimits_{t^{\prime}=1}^T{\sqrt{( 1-\psi _{t}^{2} ) ( 1-\psi _{t^{\prime}}^{2} )}}}\\
		&\times h_{\mathrm{b}}^{*}({L_x\psi _t}/{2},{L_z\psi _{t^{\prime}}}/{2})h_{\mathrm{e}}({L_x\psi _t}/{2},{L_z\psi _{t^{\prime}}}/{2}),
	\end{split}	
\end{align}
where $T$ is a complexity-vs-accuracy tradeoff parameter, and $\psi _t=\cos \left( \frac{\left( 2t-1 \right) \pi}{2T} \right) $.
\end{lemma}
\vspace{-2pt}
\begin{IEEEproof}
Please refer to Appendix \ref{Appendix:G} for more details.
\end{IEEEproof}
Given that the MSR in \eqref{max_secrecy_rate} is expressed as a function of the channel gain of each user and the channel correlation factor, after deriving $g_{k}^{\mathrm{c}}$ and $\rho _{\mathrm{c}}$, we can obtain the MSR achieved by the planar CAPA as $\mathcal{R}_{\mathrm{c}}=R\left( g_{\mathrm{b}}^{\mathrm{c}},g_{\mathrm{e}}^{\mathrm{c}},\overline{\rho} _{\mathrm{c}} \right) $, where $\overline{\rho }_{\mathrm{c}}=\frac{\left| \rho_{\mathrm{c}} \right|^2}{g_{\mathrm{b}}^{\mathrm{c}}g_{\mathrm{e}}^{\mathrm{c}}}$.

To further elucidate the properties of planar CAPAs in the secure transmission, we analyze the asymptotic MSR where $\mathcal{A}$ becomes infinitely large, i.e., $L_x,L_z\rightarrow\infty$. Under this condition, the channel gain for user $k\in\{{\mathrm{b},\mathrm{e}}\}$ satisfies
\begin{align}\label{asy_g_planarcapa}
&\lim_{L_x,L_z\rightarrow \infty} g_{k}^{\mathrm{c}}\notag\\
&=\lim_{L_x,L_z\rightarrow \infty} \frac{k_{0}^{2}\eta ^2}{4\pi}\sum_{x\in \mathcal{X} _k}{\sum_{z\in \mathcal{Z} _k}{\arctan \left( \frac{xz}{\Psi _k\sqrt{\Psi _{k}^{2}+x^2+z^2}} \right)}}\notag\\
&=\frac{k_{0}^{2}\eta ^2}{4\pi}\times 4\arctan \left( \lim_{x,z\rightarrow \infty} \frac{xz}{\Psi _k\sqrt{\Psi _{k}^{2}+x^2+z^2}} \right) \notag\\
&=\frac{k_{0}^{2}\eta ^2}{4\pi}\times 4\arctan \left( \infty \right) =\frac{k_{0}^{2}\eta ^2}{4\pi}\times 4\frac{\pi}{2}=\frac{k_{0}^{2}\eta ^2}{2},
\end{align}
which is a constant value. 

On the other hand, characterizing $\lim_{L_x,L_z\rightarrow\infty}\overline{\rho }_{\mathrm{c}}$ is more involved. Intuitively, as the aperture of the CAPA becomes infinitely large, the spatial resolution of the array increases, and the overlap between the effective channel patterns of Bob and Eve becomes negligible. This results in an increasingly orthogonal spatial response, hence a diminishing correlation \cite{dai}. Importantly, although one might expect the channel correlation to vanish in the infinite-aperture limit, recent work \cite{correlation} applies the stationary phase method to derive an expression for $\lim_{L_x,L_z\rightarrow\infty}\overline{\rho }_{\mathrm{c}}$, which shows that it converges to a non-zero constant that is nevertheless much smaller than one, i.e., $1-\lim_{L_x,L_z\rightarrow\infty}\overline{\rho }_{\mathrm{c}}\approx 1$. Owing to the complexity of the resulting expression, we omit it here; the numerical results reported in \cite{boqun_jstsp} are consistent with this behavior. Consequently, based on these findings, we determine the asymptotic MSR as follows.
\vspace{-5pt}
\begin{corollary}\label{cor_planar_capa}
When $L_x,L_z\rightarrow\infty$, the asymptotic MSR for the planar CAPA is given by
\begin{equation}\label{asyR_planar_capa}
	\lim_{L_x,L_z\rightarrow \infty} \mathcal{R} _{\mathrm{c}}\approx \log _2\left(1+\frac{\overline{\gamma }_{\mathrm{b}}k_{0}^{2}\eta ^2}{2}\right).	
\end{equation}
\end{corollary}
\vspace{-2pt}
\begin{IEEEproof}
The results are obtained by substituting $1-\overline{\rho }\approx 1$ and the results of \eqref{asy_g_planarcapa} into \eqref{max_secrecy_rate}.
\end{IEEEproof}
\vspace{-3pt}
\begin{remark}\label{rem_finite}
	The MSR will converge to a finite value rather than increasing indefinitely with the aperture size. This behavior is physically reasonable, as it aligns with the principle of energy conservation---the total radiated power is bounded and does not scale with the array size.
\end{remark}
\vspace{-5pt}
\begin{remark}\label{rem_current}
For the infinitely large planar CAPA, the received power at Bob is maximized by the MRT source current, i.e., $j_{\mathsf{msr}}\left( \mathbf{s} \right) =j_{\mathsf{mrt}}\left( \mathbf{s} \right)$, while Eve's received power is minimized to zero, reaching the upper limit of secrecy performance.
\end{remark}

\subsubsection{Planar SPDA}
\begin{figure}[!t]
	\centering
	\includegraphics[height=0.23\textwidth]{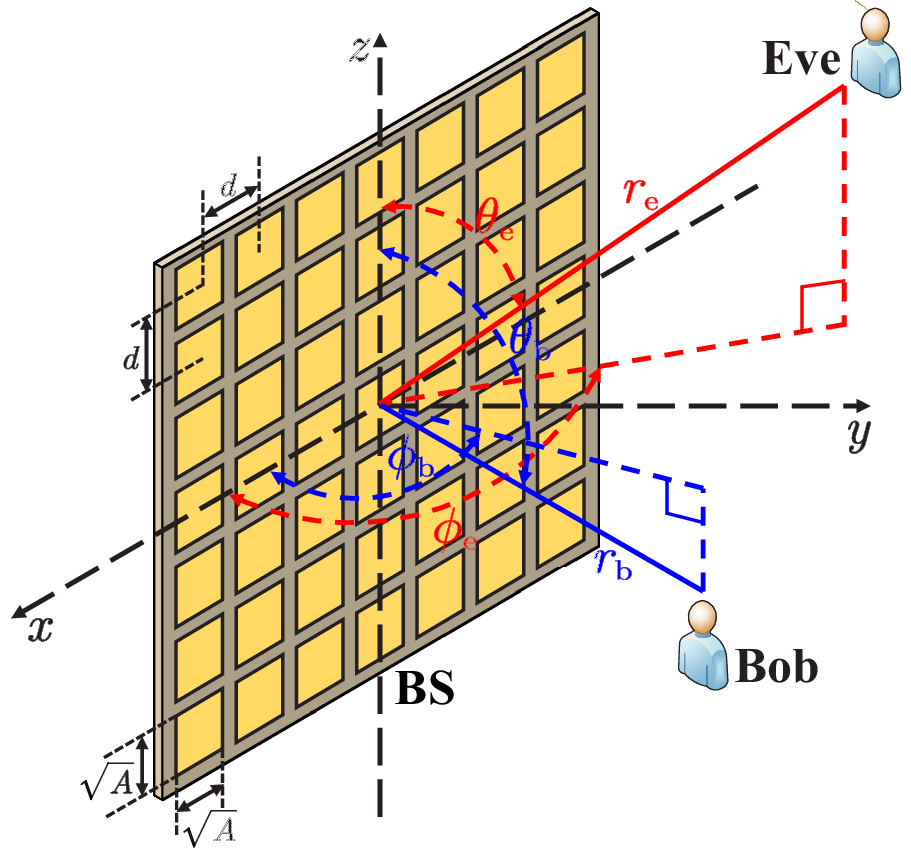}
	\caption{Illustration for Planar SPDA.}
	\vspace{-7pt}
	\label{planar_spda}
\end{figure}

For comparison, we next examine a case where the above planar CAPA is partitioned into $M=M_zM_x$ spatially discrete elements. Here, $M_{x}=2\tilde{M}_x+1$ and $M_{z}=2\tilde{M}_z+1$ represent the number of antenna elements along the $x$- and $z$-axes respectively, as illustrated in Fig. \ref{planar_spda}. The size of each element is denoted by $\sqrt{{A}_{\mathrm{s}}}\times\sqrt{{A}_{\mathrm{s}}}$, and the inter-element distance is $d$, where $\sqrt{{A}_{\mathrm{s}}}\leq d\ll r_k$. Under this configuration, the central location of each element is $[m_xd,0,m_zd]^{\mathsf{T}}$, where $m_x\in \mathcal{M} _x\triangleq\{0,\pm1,\ldots,\pm\tilde{M}_x\}$ and $m_z\in \mathcal{M} _z\triangleq\{0,\pm1,\ldots,\pm\tilde{M}_z\}$. Additionally, we have $L_x\approx M_xd$, $L_z\approx M_zd$, and 
\begin{equation}
	\mathcal{A} =\left\{ (m_xd+\ell _x,0,m_zd+\ell _z)\left| \begin{array}{c}
		\ell _x,\ell _z\in [-\frac{\sqrt{A_{\mathrm{s}}}}{2},\frac{\sqrt{A_{\mathrm{s}}}}{2}],\\
		m_x\in \mathcal{M} _x,m_z\in \mathcal{M} _z\\
	\end{array} \right. \right\}.
\end{equation}

For the planar SPDA, given that the size of each element is much smaller than the distance between the BS and the user, i.e.,$\sqrt{{A}_{\mathrm{s}}}\ll r_k$, the variation in the channel across an element is negligible. Consequently, we can express the channel gain and correlation factor as follows:
\begin{align}
	g_{k}^{\mathrm{s}}&=A_{\mathrm{s}}\sum_{m_x\in \mathcal{M} _x}^{}{\sum_{m_z\in \mathcal{M} _z}^{}{\left| h_k(m_xd,m_zd) \right|}^2},\label{g_planar_spda0}\\
	\rho _{\mathrm{s}}&=A_{\mathrm{s}}\sum_{m_x\in \mathcal{M} _x}{\sum_{m_z\in \mathcal{M} _z}{h_{\mathrm{b}}^{*}(m_xd,m_zd)h_{\mathrm{e}}(m_xd,m_zd)}}.	
\end{align}
Notably, the channel gain $g_{k}^{\mathrm{s}}$ can be calculated as follow.
\vspace{-3pt}
\begin{lemma}\label{lem_planar_spda}
	With the planar SPDA, the channel gain for user $k\in{\mathrm{b},\mathrm{e}}$ is given by
	\begin{equation}\label{g_planar_spda}
	 g_k^{\mathrm{s}}=\!\frac{A_{\mathrm{s}}k_{0}^{2}\eta ^2}{4\pi d^2}\!\sum_{x\in \mathcal{X} _k}{\sum_{z\in \mathcal{Z} _k}\!{\arctan}}\bigg( \frac{xz}{\Psi _k\sqrt{\Psi _{k}^{2}\!+\!x^2\!+\!z^2}} \bigg)\!=\zeta _{\mathrm{oc}}g_k^{\mathrm{c}},
	\end{equation}
	where $\zeta _{\mathrm{oc}}\triangleq\frac{A_{\mathrm{s}}}{d^2}$ represents the array occupation ratio (AOR).
\end{lemma}
\vspace{-5pt}
\begin{IEEEproof}
	Please refer to Appendix \ref{Appendix:I} for more details.
\end{IEEEproof}
\vspace{-3pt}
\begin{remark}\label{rem_aor}
It can be observed from \eqref{g_planar_spda} that the channel gain for each user achieved by the SPDA converges to that of the CAPA when $\zeta _{\mathrm{oc}}=1$. This result is intuitive, as an SPDA effectively becomes a CAPA when the AOR equals $1$.
\end{remark}
\vspace{-3pt}
Having derive the MSR for the SPDA, viz., $\mathcal{R}_{\mathrm{s}}=R\left( g_{\mathrm{b}}^{\mathrm{s}},g_{\mathrm{e}}^{\mathrm{s}},\overline{\rho} _{\mathrm{s}} \right) $ with $\overline{\rho }_{\mathrm{s}}=\frac{\left| \rho_{\mathrm{s}} \right|^2}{g_{\mathrm{b}}^{\mathrm{s}}g_{\mathrm{e}}^{\mathrm{s}}}$, we can then obtain its asymptotic expression by following the steps similar to the derivations for the planar CAPA. In particular, as the number of antenna element $M_x,M_z$ grows, the MSR achieved by the SPDA approaches to the following upper bound:
\begin{equation}\label{asy_SPDA}
\lim_{M_x,M_z\rightarrow \infty} \mathcal{R} _{\mathrm{s}}\approx \log _2\left( 1+\frac{\overline{\gamma }_{\mathrm{b}}k_{0}^{2}\eta ^2\zeta _{\mathrm{oc}}}{2} \right) .	
\end{equation}
\section{Analysis of the Minimum Required Power}\label{sec_MRP}
In this section, we will analyze the MRP. 
\subsection{Problem Reformulation}
The problem in \eqref{Minimum_Power_Definition} can be rewritten as follows:
\begin{subequations}
\begin{align}
	\min_{{p},\kappa (\mathbf{s})}~~ &~p\\
	{\rm{s.t.}}~~~&\frac{1+p\left| \mathcal{A} _{\mathrm{b}} \right|\sigma _{\mathrm{b}}^{-2}\left| \int_{\mathcal{A}}{h_{\mathrm{b}}}(\mathbf{s})\kappa (\mathbf{s})\mathrm{d}\mathbf{s} \right|^2}{1+p\left| \mathcal{A} _{\mathrm{e}} \right|\sigma _{\mathrm{e}}^{-2}\left| \int_{\mathcal{A}}{h_{\mathrm{e}}}(\mathbf{s})\kappa (\mathbf{s})\mathrm{d}\mathbf{s} \right|^2}\geq2^{\mathsf{R}_0},\label{constraint_1}\\
	&p> 0,~\int_{\mathcal{A}}{\left| \kappa (\mathbf{s}) \right|^2\mathrm{d}\mathbf{s}}=1,
\end{align}
\end{subequations}
where $\int_{\mathcal{A}}{\left| j(\mathbf{s}) \right|^2\mathrm{d}\mathbf{s}}= p$ and $\kappa (\mathbf{s})= \frac{j(\mathbf{s})}{\sqrt{p}}$. After multiplying both sides of \eqref{constraint_1} by $1+p\left| \mathcal{A} _{\mathrm{e}} \right|\sigma _{\mathrm{e}}^{-2}\left| \int_{\mathcal{A}}{h_{\mathrm{e}}}(\mathbf{s})\kappa (\mathbf{s})\mathrm{d}\mathbf{s} \right|^2$ and performing some basic manipulations, we obtain
\begin{equation}
	\int_{\mathcal{A}}{\int_{\mathcal{A}}{}p\kappa ^*(\mathbf{s})D\left( \mathbf{s},\mathbf{s}^{\prime} \right) \kappa (\mathbf{s}^{\prime})\mathrm{d}\mathbf{s}}\mathrm{d}\mathbf{s}^{\prime}\ge 2^{\mathsf{R}_0}-1,
\end{equation}
where
\begin{equation}\label{mrp_problem_1}
	D\left( \mathbf{s},\mathbf{s}^{\prime} \right) =\frac{\left| \mathcal{A} _{\mathrm{b}} \right|}{\sigma _{\mathrm{b}}^{2}}h_{\mathrm{b}}(\mathbf{s})h_{\mathrm{b}}^{*}(\mathbf{s}^{\prime})-2^{\mathsf{R}_0}\frac{\left| \mathcal{A} _{\mathrm{e}} \right|}{\sigma _{\mathrm{e}}^{2}}h_{\mathrm{e}}(\mathbf{s})h_{\mathrm{e}}^{*}(\mathbf{s}^{\prime}).
\end{equation}
Furthermore, we have
\begin{equation}\label{mrp_problem_2}
\int_{\mathcal{A}}{\int_{\mathcal{A}}{}p\kappa ^*(\mathbf{s})D\left( \mathbf{s},\mathbf{s}^{\prime} \right) \kappa (\mathbf{s}^{\prime})\mathrm{d}\mathbf{s}}\mathrm{d}\mathbf{s}^{\prime}\le p\lambda _{D}^{\max},	
\end{equation}
where $\lambda _{D}^{\max}$ denotes the principal eigenvalue of operator $D\left( \mathbf{s},\mathbf{s}^{\prime} \right)$. By combining \eqref{mrp_problem_1} with \eqref{mrp_problem_2}, it follows that $p\geq\frac{2^{\mathsf{R}_0}-1}{\lambda _{D}^{\max}}>0$. This implies that the minimum value of $p$, i.e., the MRP, can be expressed as follows:
\begin{equation}
	\mathcal{P}_{\star}=\frac{2^{\mathsf{R}_0}-1}{\lambda _{D}^{\max}}.
\end{equation}
In this case, we have
\begin{equation}
\int_{\mathcal{A}}{\int_{\mathcal{A}}\kappa ^*(\mathbf{s})D\left( \mathbf{s},\mathbf{s}^{\prime} \right) \kappa (\mathbf{s}^{\prime})\mathrm{d}\mathbf{s}}\mathrm{d}\mathbf{s}^{\prime}=\lambda _{D}^{\max},
\end{equation}
which indicates that the associate $\kappa(\mathbf{s})$ aligns with the normalized principal eigenfunction of $D\left( \mathbf{s},\mathbf{s}^{\prime} \right)$.
\subsection{Minimum Required Power \& Optimal Current Distribution}
The above arguments imply that the MRP and the associate current distribution correspond to the principal eigenvalue and eigenfunction of $D\left( \mathbf{s},\mathbf{s}^{\prime} \right)$, respectively. Their closed-form expressions are given as follows.
\vspace{-5pt}
\begin{theorem}
The MRP for CAPA-based secure transmission to guarantee a target secrecy rate $\mathsf{R}_0$ can be expressed as follows:
\begin{equation}\label{MRP_expression}
\mathcal{P}_{\star} =\frac{2\left( 2^{\mathsf{R}_0}-1 \right)}{\alpha +\sqrt{\alpha ^{2}+\beta }},	
\end{equation}
where $\alpha =\left| \mathcal{A} _{\mathrm{b}} \right|\sigma _{\mathrm{b}}^{-2}g_{\mathrm{b}}-2^{\mathsf{R}_0}\left| \mathcal{A} _{\mathrm{e}} \right|\sigma _{\mathrm{e}}^{-2}g_{\mathrm{e}}$, and $\beta =2^{\mathsf{R}_0+2}\left| \mathcal{A} _{\mathrm{b}} \right|\left| \mathcal{A} _{\mathrm{e}} \right|\sigma _{\mathrm{b}}^{-2}\sigma _{\mathrm{e}}^{-2}g_{\mathrm{b}}g_{\mathrm{e}}\left( 1-\overline{\rho } \right) $. The optimal current distribution is given by
\begin{equation}
	\begin{split}
j_{\mathsf{mrp}}\left( \mathbf{s} \right) =\sqrt{\mathcal{P}_{\star}}\frac{h^*_{\mathrm{b}}(\mathbf{s})-\tau h^*_{\mathrm{e}}(\mathbf{s})}
{\sqrt{\int_{\mathcal{A}}\left\lvert h^*_{\mathrm{b}}(\mathbf{s})-\tau h^*_{\mathrm{e}}(\mathbf{s})\right\rvert^2{\rm{d}}{\mathbf{s}}}},
	\end{split}
\end{equation}
where $\tau=\frac{{\left| \mathcal{A} _{\mathrm{b}} \right|g_{\mathrm{b}}}{\sigma _{\mathrm{b}}^{-2}}+{\left| \mathcal{A} _{\mathrm{e}} \right|g_{\mathrm{e}}2^{\mathsf{R}_0}}{\sigma _{\mathrm{e}}^{-2}}-\sqrt{\alpha ^{2}+\beta}}{2\left| \mathcal{A} _{\mathrm{b}} \right|\sigma _{\mathrm{b}}^{-2}\rho }$.
\end{theorem}
\vspace{-5pt}
\begin{IEEEproof}
The proof is similar to that of \textbf{Lemma~\ref{lem_eigenvalue}}. 
\end{IEEEproof}
\vspace{-5pt}
\begin{remark}\label{rem_gen2}
	The MRP given in \eqref{MRP_expression} is determined by the channel gain for each user and the channel correlation factor. This expression is applicable to any aperture, regardless of its location, shape, and size. 
\end{remark}
\vspace{-5pt}
We next compare the MRP with the required power achieved by ZF and MRT beamforming. When MRT beamforming is utilized, we have $j({\mathbf{s}})=\sqrt{p}\frac{h^*_{\mathrm{b}}(\mathbf{s})}{{\sqrt{\int_{\mathcal{A}}\left\lvert h^*_{\mathrm{b}}(\mathbf{s})\right\rvert^2{\rm{d}}{\mathbf{s}}}}}$ (as per \eqref{Optimal_Current_Rate_Low_SNR}), and the minimum required transmission power to guarantee a secrecy transmission rate of $\mathsf{R}_0$ is given by 
\begin{equation}\label{P_mrt}
\mathcal{P} _{\mathsf{mrt}}=\frac{2^{\mathsf{R}_0}-1}{\left| \mathcal{A} _{\mathrm{b}} \right|\sigma _{\mathrm{b}}^{-2}g_{\mathrm{b}}-2^{\mathsf{R}_0}\left| \mathcal{A} _{\mathrm{e}} \right|\sigma _{\mathrm{e}}^{-2}g_{\mathrm{e}}\overline{\rho }}.
\end{equation}
By noting that fact that $\overline{\rho }\in[0,1]$, we can prove that 
\begin{align}
\frac{\mathcal{P} _{\mathsf{mrt}}}{\mathcal{P}_{\star}}=\frac{1}{2}\frac{\alpha +\sqrt{\alpha ^{2}+\beta }}{\left| \mathcal{A} _{\mathrm{b}} \right|\sigma _{\mathrm{b}}^{-2}g_{\mathrm{b}}-2^{\mathsf{R}_0}\left| \mathcal{A} _{\mathrm{e}} \right|\sigma _{\mathrm{e}}^{-2}g_{\mathrm{e}}\overline{\rho }}>1.
\end{align}
Furthermore, we have the following observations.
\vspace{-5pt}
\begin{remark}\label{rem_MRP}
It is observed that the proposed optimal current distribution outperforms MRT beamforming in terms of minimizing the transmit power. Besides, the optimal current distribution $j_{\mathsf{mrp}}\left( \mathbf{s} \right) $ can achieve an arbitrary large target secrecy rate as long as the transmission power is sufficiently high. In contrast, the MRT cannot achieve a secrecy rate greater than $\log _2\left( \left| \mathcal{A} _{\mathrm{b}} \right|\sigma _{\mathrm{b}}^{-2}g_{\mathrm{b}} \right) -\log _2\left( \left| \mathcal{A} _{\mathrm{e}} \right|\sigma _{\mathrm{e}}^{-2}g_{\mathrm{e}} \overline{\rho }\right)$, regardless of the transmission power level. 
\end{remark}
\vspace{-5pt}
We then consider ZF beamforming, which yields $\sqrt{p}\frac{h^*_{\mathrm{b}}(\mathbf{s})-\frac{g_{\mathrm{b}}}{\rho}\overline{\rho }h^*_{\mathrm{e}}(\mathbf{s})}
{\sqrt{\int_{\mathcal{A}}\left\lvert h^*_{\mathrm{b}}(\mathbf{s})-\frac{g_{\mathrm{b}}}{\rho}\overline{\rho }h^*_{\mathrm{e}}(\mathbf{s})\right\rvert^2{\rm{d}}{\mathbf{s}}}}$ (as per \eqref{Optimal_Current_Rate_High_SNR}), and the minimum required transmission power to guarantee a secrecy transmission rate of $\mathsf{R}_0$ is given by 
\begin{equation}\label{P_zf}
\mathcal{P} _{\mathsf{zf}}=\frac{2^{\mathsf{R}_0}}{\left| \mathcal{A} _{\mathrm{b}} \right|\sigma _{\mathrm{b}}^{-2}g_{\mathrm{b}}\left( 1-\overline{\rho } \right)}.
\end{equation}
By letting ${\mathsf{R}_0}\rightarrow\infty$, we have
\begin{align}
\lim_{{\mathsf{R}_0}\rightarrow\infty}\frac{1}{\alpha +\sqrt{\alpha ^{2}+\beta }}
&=\lim_{{\mathsf{R}_0}\rightarrow\infty}\frac{\sqrt{\alpha ^{2}+\beta }-\alpha}{\beta}\notag\\
&=\frac{1}{2\left| \mathcal{A} _{\mathrm{b}} \right|\sigma _{\mathrm{b}}^{-2}g_{\mathrm{b}}\left( 1-\overline{\rho } \right)},
\end{align}
which suggests that
\begin{align}\label{rem_MRP_zf_compare}
\lim_{{\mathsf{R}_0}\rightarrow\infty}\mathcal{P}_{\star}\simeq\frac{2^{\mathsf{R}_0}}{\left| \mathcal{A} _{\mathrm{b}} \right|\sigma _{\mathrm{b}}^{-2}g_{\mathrm{b}}\left( 1-\overline{\rho } \right)}
=\mathcal{P} _{\mathsf{zf}}.
\end{align}
\vspace{-5pt}
\begin{remark}\label{rem_MRP_zf}
The results in \eqref{rem_MRP_zf_compare} suggest that ZF beamforming is asymptotically optimal when achieving an infinitely large secrecy rate target.
\end{remark}
\vspace{-5pt}
\subsection{Typical Cases}
Since the MRP is expressed as a function of the channel gains and the channel correlation factor, the MRP for both the planar CAPA and SPDA under the LoS channel can be readily obtained based on the results in \textbf{Lemma~\ref{lem_planar_capa}} and \textbf{\ref{lem_planar_spda}}, respectively. Therefore, in this subsection, we will focus on the asymptotic behaviors of $\mathcal{P}$ as $L_x,L_z\rightarrow\infty$ or $M_x,M_z\rightarrow\infty$ to gain further insights, where the considered channel model and array structure are the same as those detailed in Section \ref{discussion_MSR}.
\subsubsection{Planar CAPA}
The asymptotic MRP for the planar CAPA with infinitely large aperture is given as follows.
\vspace{-5pt}
\begin{corollary}\label{cor_MRP_capa}
When $L_x,L_z\rightarrow\infty$, the MRP satisfies
\begin{equation}
\lim_{L_x,L_z\rightarrow \infty} \mathcal{P}_{\star} \approx \frac{2\left( 2^{\mathsf{R}_0}-1 \right)}{k_{0}^{2}\eta ^2\left| \mathcal{A} _{\mathrm{b}} \right|\sigma _{\mathrm{b}}^{-2}}.
\end{equation}
\end{corollary}
\vspace{-3pt}
\begin{IEEEproof}
	Similar to the proof of \textbf{Corollary~\ref{cor_planar_capa}}.
\end{IEEEproof}
\vspace{-5pt}
\begin{remark}\label{rem_lowerbound}
The results of \textbf{Corollary~\ref{cor_MRP_capa}} suggest that as the aperture size increases, the MRP for the planar CAPA will converge to a constant that is larger than zero.
\end{remark}
\vspace{-5pt}
\subsubsection{Planar SPDA}
In the case of the planar SPDA, when the antenna number grows to infinity, i.e., $M_x,M_z\rightarrow\infty$, the MRP approaches to the follow lower bound:
\begin{equation}
\lim_{M_x,M_z\rightarrow \infty} \mathcal{P}_{\star}  \approx \frac{2\left( 2^{\mathsf{R}_0}-1 \right)}{k_{0}^{2}\eta ^2\left| \mathcal{A} _{\mathrm{b}} \right|\sigma _{\mathrm{b}}^{-2}\zeta _{\mathrm{oc}}}.
\end{equation}
We note that for extremely large arrays, the minimum transmit power required by an SPDA for a given target secrecy rate reduces to that of the CAPA when $\zeta _{\mathrm{oc}}=1$.

	\begin{figure}[!t]
	\centering
	\subfigbottomskip=3pt
	\subfigcapskip=-2pt
	\setlength{\abovecaptionskip}{3pt}   
	\subfigure[$\overline{\rho}$ vs. $T$.]
	{
		\includegraphics[height=0.13\textwidth]{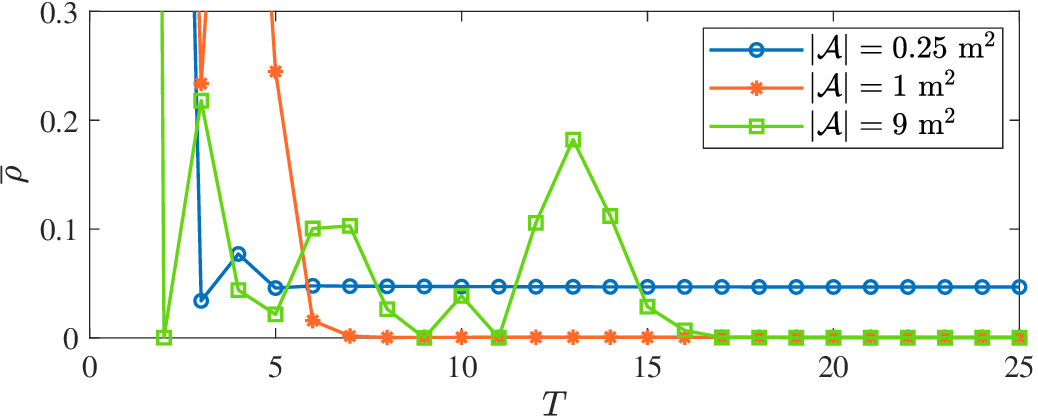}
		\label{rho1}	
	}\vspace{-3pt}	
	\subfigure[$\overline{\rho}$ vs. $T$, $\lvert \mathcal{A}\rvert=10^6$ m$^2$.]
	{
		\includegraphics[height=0.13\textwidth]{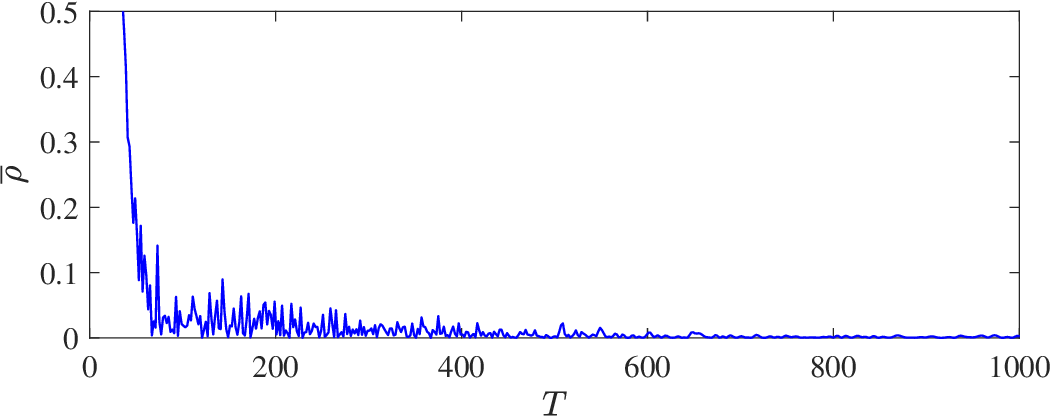}
		\label{rho2}	
	}
	\caption{Convergence of Chebyshev-Gauss quadrature.}
	\label{Convergence}
	\vspace{-5pt}
\end{figure}

\section{Numerical Results}\label{numerical}
In this section, numerical results are presented to demonstrate the secrecy performance achieved by CAPAs. For clarity, the simulations employ the LoS model and planar arrays specified in Section \ref{discussion_MSR}. Without otherwise specification, the simulation parameters are set as follows: $\lambda=0.125$ m, $\left| \mathcal{A} _{\mathrm{b}} \right|=\left| \mathcal{A} _{\mathrm{e}} \right|=\frac{\lambda^2}{4\pi}$, $P=30$ dB, ${\sigma^2_{\mathrm{b}}}={\sigma^2_{\mathrm{e}}}=20$ dB, $L_x=L_z$, $(r_{\mathrm{b}},\theta_{\mathrm{b}},\phi_{\mathrm{b}})=(10 \, \rm{m},\frac{\pi}{6},\frac{\pi}{6})$, and $(r_{\mathrm{e}},\theta_{\mathrm{e}},\phi_{\mathrm{e}})=(20 \, \rm{m},\frac{\pi}{3},\frac{\pi}{3})$.

In order to provide a principled guideline for selecting the accuracy–complexity tradeoff parameter $T$, we first examine the convergence of $\overline{\rho}$ as a function of $T$ for the aperture sizes considered in the subsequent simulations, as shown in Fig. \ref{rho1}. For all aperture sizes in Fig. \ref{rho1}, $\overline{\rho}$
stabilizes when $T\geq20$, indicating that $T=20$ Chebyshev–Gauss nodes are sufficient to accurately approximate the channel correlation factor. We also note that larger apertures require larger $T$ to reach convergence. To ensure reliable convergence for extremely large arrays, we further consider an aperture size of $10^6$ m$^2$ in Fig. \ref{rho2}, where $\overline{\rho}$ effectively converges at approximately $T\approx700$. Guided by these results and to balance computational cost and accuracy, we set $T=700$ in Fig. \ref{R_size} and \ref{P_size}, which study asymptotic behavior for extremely large arrays, and use $T=20$ for all other results.

It is worth noting that, under our simulation settings, the online computation of $g_k$ admits the closed-form expression in \eqref{g_planar_capa} and can therefore be evaluated efficiently in constant time, independent of the aperture size. The channel correlation $\rho$ is evaluated via the Chebyshev–Gauss quadrature in \eqref{rho_planar_capa}, with complexity $\mathcal{O}(T^2)$. As noted earlier, larger apertures generally require larger $T$ for convergence; however, Fig. \eqref{rho1} indicates that a modest fixed value (e.g., $T=100$) is sufficient for most practical scenarios. Consequently, the computation of $\rho$ can also be treated as effectively independent of the aperture size.

\begin{figure}[!t]
	\centering
	\setlength{\abovecaptionskip}{3pt}
	\includegraphics[height=0.25\textwidth]{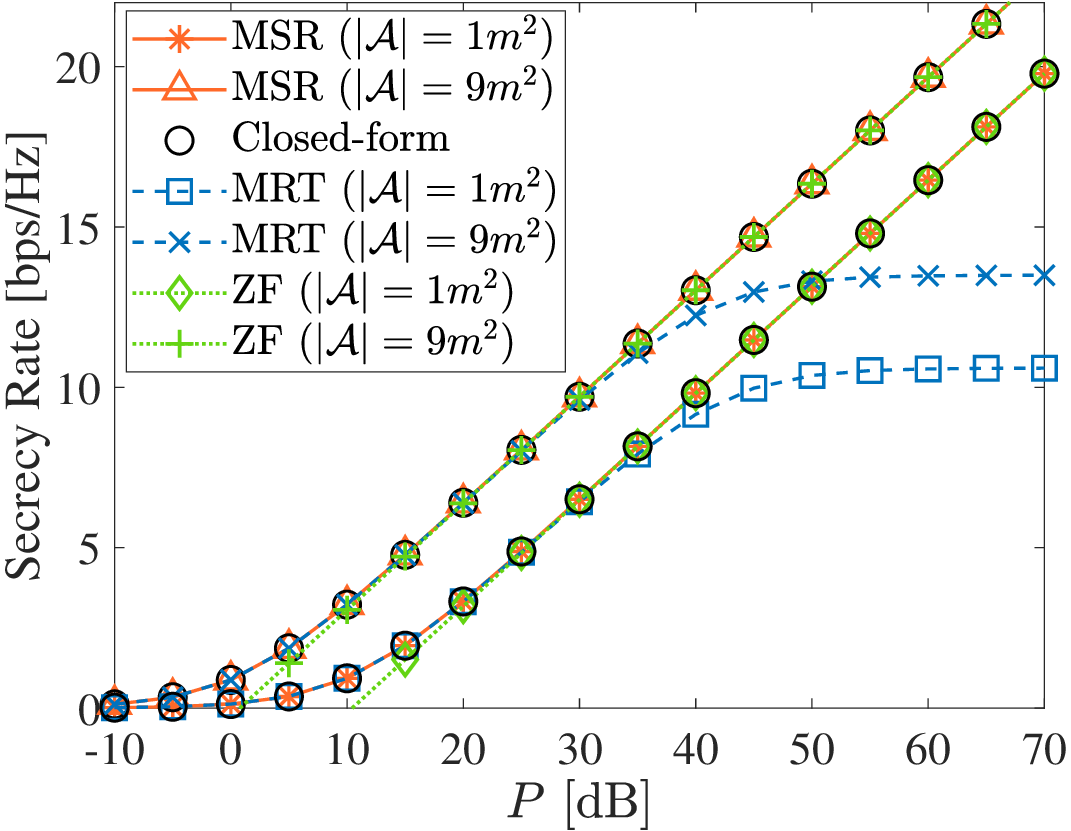}
	\caption{Secrecy rates versus power budget $P$.}
	\vspace{-7pt}
	\label{R_P}
\end{figure}

\subsection{Maximum Secrecy Rate}
{\figurename} {\ref{R_P}} illustrates the secrecy rate achieved by the CAPA with different aperture sizes as a function of the power budget. As observed, the derived closed-form expressions closely align with the simulation results, which validates the correctness of our previously derived results. For comparison, the secrecy rates achieved by MRT and ZF beamforming schemes are also presented. The results show that the proposed optimal current distribution yields better secrecy rate than both the MRT and ZF-based schemes. Moreover, as the transmit power increases, the secrecy rate achieved by MRT beamforming converges to a finite constant, while the rates achieved by ZF beamforming and the optimal current distribution increase monotonically. This demonstrates that the high-SNR slope for the MSR is greater than that achieved by MRT beamforming.

From {\figurename} {\ref{R_P}}, we can also observe that in the low-SNR regime, the secrecy rate achieved by MRT beamforming is nearly identical to that achieved by the optimal current distribution. This is because, at low SNRs, Gaussian noise dominates over information leakage, making MRC beamforming more effective for enhancing the secrecy rate. In contrast, at high SNRs, the secrecy rate achieved by ZF beamforming closely approaches that of the optimal current distribution. This is because information leakage becomes the dominant factor over Gaussian noise in the high-SNR regime, making leakage cancellation essential for improving the secrecy rate. These observations align with the discussions in Section \ref{Section: Asymptotic Analysis}.

\begin{figure}[!t]
	\centering
	\setlength{\abovecaptionskip}{3pt}
	\includegraphics[height=0.25\textwidth]{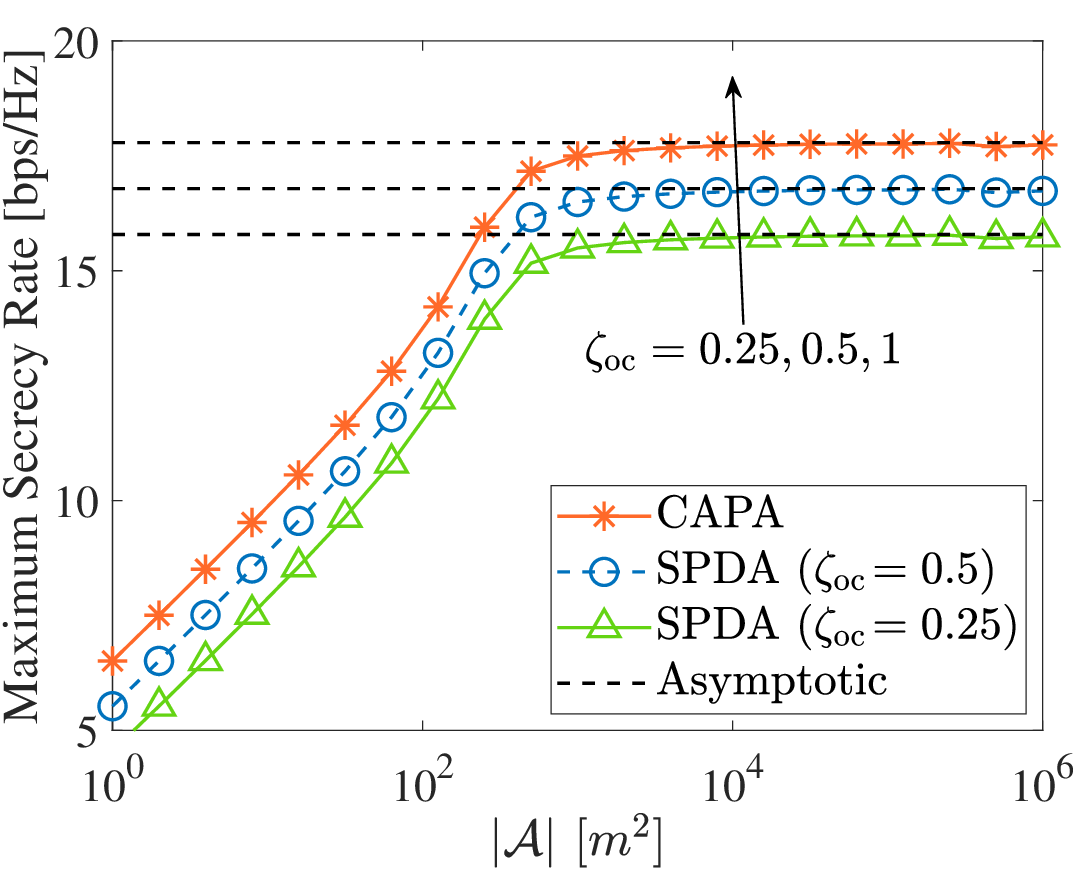}
	\caption{MSRs versus aperture size $\left| \mathcal{A}\right|$.}
	\vspace{-7pt}
	\label{R_size}
\end{figure}

\begin{figure}[!t]
	\centering
	\setlength{\abovecaptionskip}{3pt}
	\includegraphics[height=0.25\textwidth]{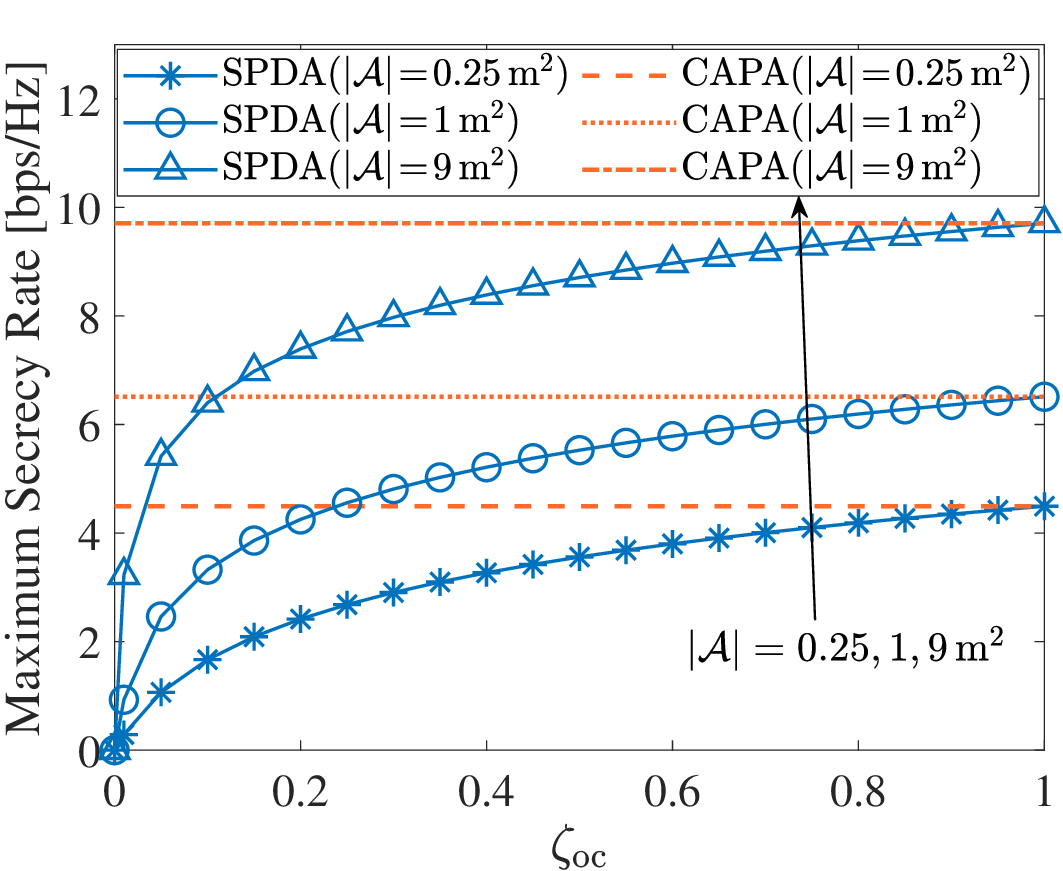}
	\caption{{MSRs versus AOR $\zeta_{\mathrm{oc}}$.}}
	\vspace{-7pt}
	\label{R_aor}
\end{figure}

To further investigate the behavior of the MSR as the BS aperture size increases, we present the MSR as a function of $\left| \mathcal{A} \right|$ for both CAPA and SPDA in {\figurename} {\ref{R_size}}. It can be observed that as the aperture size grows, the MSRs increase and eventually approach their upper bounds, as indicated in \eqref{asyR_planar_capa} and \eqref{asy_SPDA}, which adheres to the principle of energy conservation. This observation verifies our discussions in \textbf{Remark~\ref{rem_finite}}. {Notably, although a larger aperture is able to improve the secrecy performance for CAPA systems, this comes at the cost of increased implementation complexity. Therefore, the aperture size must be carefully chosen to balance performance gains with implementation feasibility.} Additionally, the results in {\figurename} {\ref{R_size}} demonstrate that for the same aperture size, CAPA achieves superior secrecy performance compared to conventional SPDA. To further highlight the performance gap between CAPA and SPDA in terms of achievable MSR, we plot the MSR for SPDAs against the AOR in {\figurename} \ref{R_aor}. This graph illustrates the MSR for an SPDA gradually converges with that of a CAPA as the AOR approaches one, which corroborates the statements in \textbf{Remark~\ref{rem_aor}}.

\begin{figure}[!t]
	\centering
	\setlength{\abovecaptionskip}{3pt}
	\includegraphics[height=0.28\textwidth]{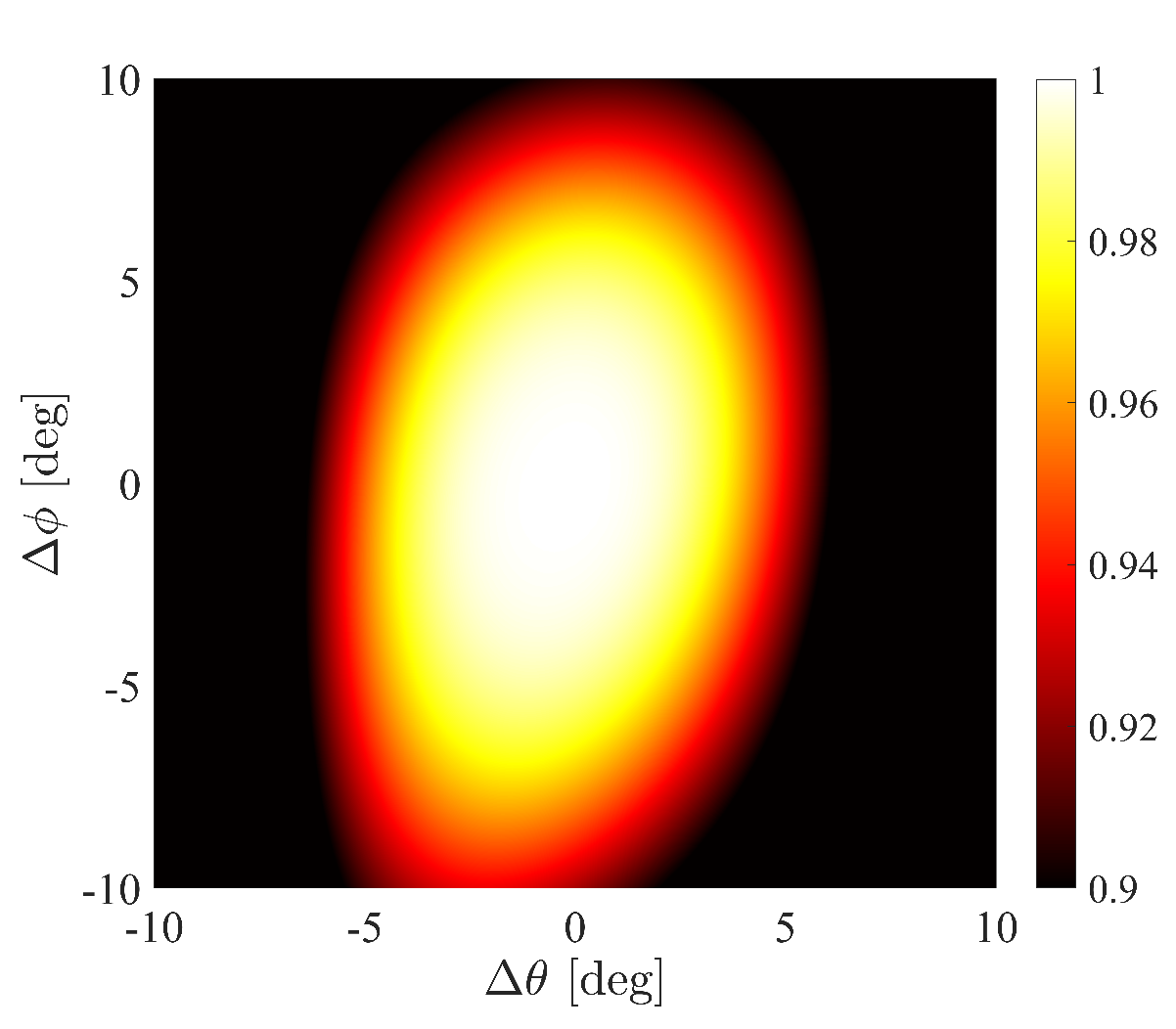}
	\caption{Normalized secrecy rate versus Eve's angular mismatch, $\lvert \mathcal{A}\rvert=0.25$ m$^2$.}
	\vspace{-10pt}
	\label{mismatch}
\end{figure}

To study the sensitivity of the proposed design to Eve's CSI errors, Fig. \ref{mismatch} depicts the normalized secrecy rate (NSR) achieved by the proposed optimal current distribution under two-dimensional angular mismatch $(\Delta\theta,\Delta\phi)$ at Eve. We observer that the NSR degrades smoothly from its peak at $(0,0)$ without an abrupt performance cliff, and remains above $0.9$ over the range $\Delta\theta\in[-5^\circ,5^\circ]$ and $\Delta\phi\in[-10^\circ,10^\circ]$. Moreover, a broad high-performance region is preserved (e.g., NSR $\geq0.95$ for several degrees of mismatch in both dimensions), demonstrating that the proposed design is robust to moderate angular mismatch in Eve’s CSI. This robustness further implies that the summary parameters $(g_\mathrm{b},g_\mathrm{e},\rho)$ do not need to be updated aggressively: as long as the accumulated angular deviation stays within the high-performance region, the resulting loss is negligible. In practice, the parameters can be updated once per coherence interval or in an event-triggered manner when the geometry/angles change appreciably (e.g., due to user motion).

\begin{figure}[!t]
	\centering
	\setlength{\abovecaptionskip}{3pt}
	\includegraphics[height=0.25\textwidth]{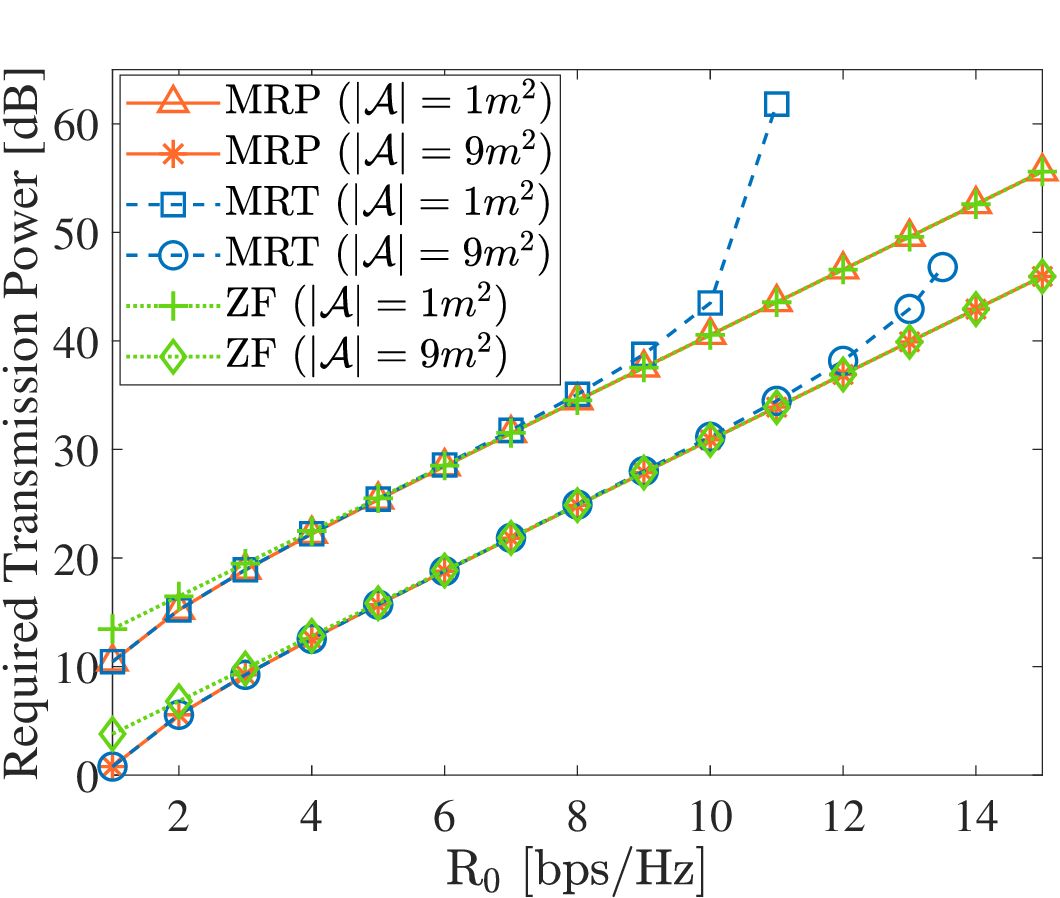}
	\caption{Required powers versus target secrecy rate $\mathsf{R}_0$.}
	\vspace{-5pt}
	\label{P_R0}
\end{figure}

\subsection{Minimum Required Power}
{\figurename} {\ref{P_R0}} illustrates the MRP and the transmission power needed for ZF and MRT beamforming to achieve different target secrecy rates ${\mathsf{R}}_0$. It can be observed that while the required powers for MRT beamforming are nearly identical to the MRP for small values of $\mathsf{R}_0$, the gap becomes significant as $\mathsf{R}_0$ grows, with MRT requiring substantially more power than the MRP. Notably, when the target secrecy rate exceeds a certain threshold, MRT beamforming becomes incapable of achieving it, regardless of the available transmission power. In contrast, the current distribution designed to achieve the MRP can support arbitrarily high secrecy rates, provided sufficient transmission power is available. These findings align with the discussions in \textbf{Remark \ref{rem_MRP}}. Furthermore, we observe that when achieving a high target secrecy rate, ZF beamforming yields performance virtually identical to that of the optimal current distribution, which corroborates the results discussed in \textbf{Remark \ref{rem_MRP_zf}}. The above observations highlight the superiority of the proposed optimal current design in terms of minimizing transmit power while guaranteeing a target secrecy rate.

\begin{figure}[!t]
	\centering
	\setlength{\abovecaptionskip}{3pt}
	\includegraphics[height=0.25\textwidth]{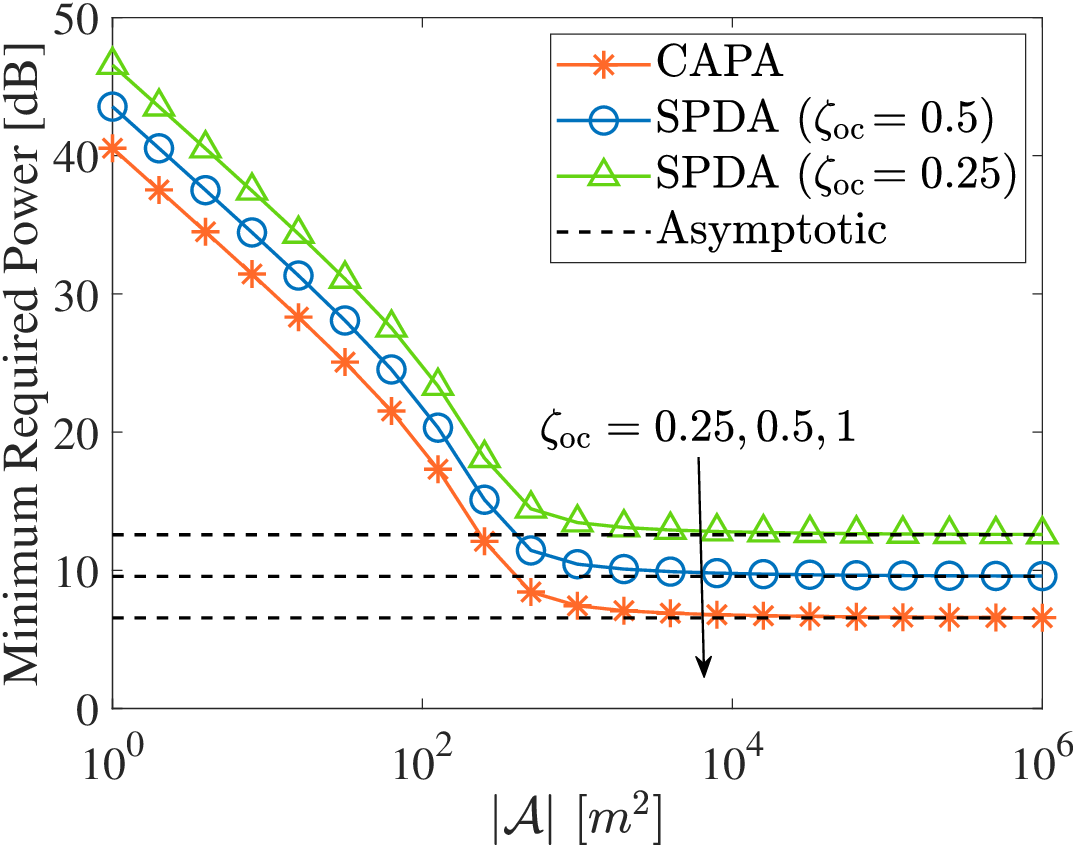}
	\caption{MRPs versus aperture size $\left| \mathcal{A}\right|$.}
	\vspace{-5pt}
	\label{P_size}
\end{figure}

\begin{figure}[!t]
	\centering
	\setlength{\abovecaptionskip}{3pt}
	\includegraphics[height=0.25\textwidth]{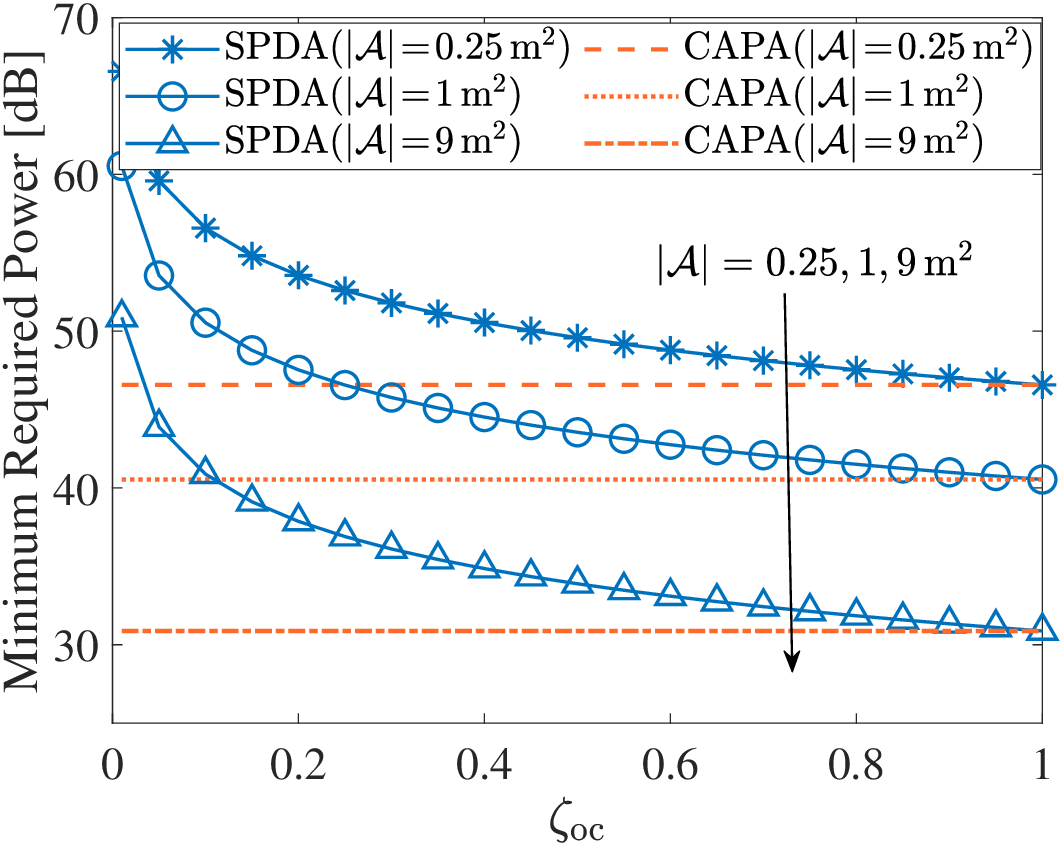}
	\caption{{MRPs versus AOR $\zeta_{\mathrm{oc}}$.}}
	\vspace{-10pt}
	\label{P_aor}
\end{figure}

In {\figurename} \ref{P_size}, we present the MRP as a function of the BS aperture size $\left| \mathcal{A} \right|$. Initially, the MRP decreases as $\left| \mathcal{A} \right|$ increases, which indicates that using a larger aperture can increase the power utilization efficiency. However, as the aperture size continues to grow, the MRP converges to a non-zero lower bound. This observation is consistent with \textbf{Remark~\ref{rem_lowerbound}} and aligns with the physical limitation that the required transmission power cannot be reduced to zero, even with an extremely large array size. Additionally, {\figurename} \ref{P_aor} shows the MRP for both CAPAs and SPDAs as a function of the AOR. Similar to the MSR, the MRP for an SPDA gradually converges to that of a CAPA as the AOR approaches $1$.

\section{Conclusion}\label{conclusion}
This article has developed a novel secure transmission framework using CAPAs and analyzed its fundamental secrecy performance limits. We derived closed-form expressions for the MSR under a power constraint and the MRP to achieve a target secrecy rate, along with the corresponding optimal continuous current distributions. We proved that the rate-optimal source current simplifies to MRT beamforming in the low-SNR region and to ZF beamforming in the high-SNR region. Additionally, we showed that the power-optimal current converges to ZF beamforming in the high-SNR regime. Through both theoretical analyses and numerical simulations, we demonstrated that CAPAs provide superior secrecy performance compared to conventional SPDAs. These findings underscore the potential of CAPAs as a promising paradigm for secure wireless communication.

\begin{appendix}
\setcounter{equation}{0}
\renewcommand\theequation{A\arabic{equation}}
\subsection{Proof of \textbf{Lemma \ref{lemma_1}}}\label{Appendix:A}
Inserting \eqref{function_Q} and \eqref{Q_invert} into the left-hand side of \eqref{Q_Inversion} gives
\begin{align}\label{invert}
&\int_{\mathcal{A}}Q\left( \mathbf{s},\mathbf{s}_1 \right)\hat{Q}\left( \mathbf{s}_1,\mathbf{s}' \right){\rm{d}}\mathbf{s}_1
=\int_{\mathcal{A}}\left( \delta (\mathbf{s}-\mathbf{s}_1)+\mu h_{\mathrm{e}}(\mathbf{s})h_{\mathrm{e}}^{*}(\mathbf{s}_1) \right)\notag\\
&\times\left( \delta (\mathbf{s}_1-\mathbf{s}')-\frac{\mu h_{\mathrm{e}}(\mathbf{s}_1)h_{\mathrm{e}}^{*}(\mathbf{s}')}{1+\mu g_{\mathrm{e}}} \right) \mathrm{d}\mathbf{s}_1
=\delta (\mathbf{s}-\mathbf{s}')\notag\\
&+\left( \mu -\frac{\mu+\mu ^2g_{\mathrm{e}}}{1+\mu g_{\mathrm{e}}} \right) h_{\mathrm{e}}(\mathbf{s})h_{\mathrm{e}}^{*}(\mathbf{s}')=\delta (\mathbf{s}-\mathbf{s}').
\end{align}
Following the same approach to obtain \eqref{invert}, we also obtain
\begin{align}
\int_{\mathcal{A}}\hat{Q}\left( \mathbf{s},\mathbf{s}_1 \right)Q\left( \mathbf{s}_1,\mathbf{s}' \right){\rm{d}}\mathbf{s}_1=\delta (\mathbf{s}-\mathbf{s}').
\end{align}
This completes the proof of \textbf{Lemma~\ref{lemma_1}}.

\subsection{Proof of \textbf{Lemma \ref{lemma_2}}}\label{Appendix:B}
Inserting \eqref{function_Q} and \eqref{Q_invert} into the left-hand side of \eqref{Indentity_Transform} gives
\begin{align}\label{Denominator_Calculation_Forward_00}
			E(\mathbf{s}_1,\mathbf{s}_1')=&\int_{\mathcal{A}}{\int_{\mathcal{A}}}(\delta (\mathbf{s}_1-\mathbf{s})+\mu h_{\mathrm{e}}(\mathbf{s}_1)h_{\mathrm{e}}^{*}(\mathbf{s}))A_{\rm{e}}(\mathbf{s},\mathbf{s}^{\prime})\nonumber\\
			&\times (\delta (\mathbf{s}^{\prime}-\mathbf{s}_{1}^{\prime})+\mu h_{\mathrm{e}}(\mathbf{s}')h_{\mathrm{e}}^{*}(\mathbf{s}_{1}^{\prime}))\mathrm{d}\mathbf{s}\mathrm{d}\mathbf{s}^{\prime}.
		\end{align}
We then calculate $E(\mathbf{s}_1,\mathbf{s}^{\prime}_1)$ as follows. According to that fact of $\int_{{\mathcal{A}}}\delta({\mathbf{x}}-{\mathbf{x}}_0)f(\mathbf{x}){\rm{d}}{\mathbf{x}}=f({\mathbf{x}}_0)$, the integral in terms of ${{\mathbf{s}}}$ involved in \eqref{Denominator_Calculation_Forward_00} can be calculated as follows:
		\begin{align}\label{Denominator_Calculation_Forward_1}
				&\int_{\mathcal{A}}(\delta (\mathbf{s}_1-\mathbf{s})\!+\!\mu h_{\mathrm{e}}(\mathbf{s}_1)h_{\mathrm{e}}^{*}(\mathbf{s}))(\delta (\mathbf{s}-\mathbf{s}^{\prime})\!+\!\overline{\gamma }_{\mathrm{e}}h_{\mathrm{e}}(\mathbf{s})h_{\mathrm{e}}^{*}(\mathbf{s}^{\prime}))\mathrm{d}\mathbf{s}\notag\\
				&=\delta (\mathbf{s}_1-\mathbf{s}^{\prime})+ (\mu+\overline{\gamma }_{\mathrm{e}}+\overline{\gamma }_{\mathrm{e}}\mu g_{\mathrm{e}})h_{\mathrm{e}}(\mathbf{s}_{1})h_{\mathrm{e}}^{*}(\mathbf{s}').		
		\end{align}
	We next calculate the integral in terms of ${{\mathbf{s}}}'$, which yields
		\begin{align}\label{Denominator_Calculation_Forward_2}
			&\int_{\mathcal{A}}\eqref{Denominator_Calculation_Forward_1}\times (\delta (\mathbf{s}^{\prime}-\mathbf{s}_{1}^{\prime})+\mu h_{\mathrm{e}}(\mathbf{s}')h_{\mathrm{e}}^{*}(\mathbf{s}_{1}^{\prime}))\mathrm{d}\mathbf{s}^{\prime}\notag\\
			&=\!\delta (\mathbf{s}_1\!-\!\mathbf{s}_1^{\prime})\!+\chi h_{\mathrm{e}}(\mathbf{s}_1)h_{\mathrm{e}}^{*}(\mathbf{s}_1^{\prime}),
		\end{align}
	where $\chi =\overline{\gamma }_{\mathrm{e}}\mu ^2g_{\mathrm{e}}^{2}+\mu ^2g_{\mathrm{e}}+2\overline{\gamma }_{\mathrm{e}}\mu g_{\mathrm{e}}+2\mu +\overline{\gamma }_{\mathrm{e}}$. Inserting $\mu =-\frac{1}{g_{\mathrm{e}}}\pm \frac{1}{g_{\mathrm{e}}\sqrt{1+\overline{\gamma }_{\mathrm{e}}g_{\mathrm{e}}}}$ into the expression of $\chi$, we can obtain $\chi=0$, which yields 
	\begin{equation}\label{function_c}
		E(\mathbf{s}_1,\mathbf{s}^{\prime}_1)=\delta (\mathbf{s}_1-\mathbf{s}_1^{\prime}),
	\end{equation}
which completes the proof of \textbf{Lemma \ref{lemma_2}}.

\subsection{Proof of \textbf{Lemma \ref{Lemma_Fredholm}}}\label{Appendix:C}
Equation \eqref{C_definition} can be written as follows:
\begin{equation}\label{Eigenvalue_Equation_Basic2}
	\begin{split} &\!C(\mathbf{s}_2,\mathbf{s}_2^{\prime})\!=\!\underset{I_1}{\underbrace{\int_{\mathcal{A}}{\int_{\mathcal{A}}}\hat{Q}(\mathbf{s}_2,\mathbf{s}_1)B(\mathbf{s}_1,\mathbf{s}_{1}^{\prime})\hat{Q}(\mathbf{s}_{1}^{\prime},\mathbf{s}_{2}^{\prime})\mathrm{d}\mathbf{s}_1\mathrm{d}\mathbf{s}_{1}^{\prime}}}\\
	&~~~-\lambda _B\underset{I_2}{\underbrace{\int_{\mathcal{A}}{\int_{\mathcal{A}}{}}\hat{Q}(\mathbf{s}_2,\mathbf{s}_1)\delta (\mathbf{s}_1-\mathbf{s}_{1}^{\prime})\hat{Q}(\mathbf{s}_{1}^{\prime},\mathbf{s}_{2}^{\prime})\mathrm{d}\mathbf{s}_1\mathrm{d}\mathbf{s}_{1}^{\prime}}}.
	\end{split}
\end{equation}
Substituting \eqref{Semi_Def_Function_Original} into $I_1$ gives
\begin{align}\label{I1}
I_1&=\int_{\mathcal{A}}{\int_{\mathcal{A}}{A_{\rm{b}}(\mathbf{s},\mathbf{s}^{\prime})}}\int_{\mathcal{A}}\hat{Q}(\mathbf{s}_2,\mathbf{s}_1)Q\left(\mathbf{s}_1,\mathbf{s}\right) \mathrm{d}\mathbf{s}_1\notag\\
&\times\int_{\mathcal{A}}{Q\left(\mathbf{s}^\prime,\mathbf{s}^\prime_1 \right) \hat{Q}(\mathbf{s}_{1}^{\prime},\mathbf{s}_{2}^{\prime})}\mathrm{d}\mathbf{s}_{1}^{\prime}\mathrm{d}\mathbf{s}\mathrm{d}\mathbf{s}^{\prime}=\int_{\mathcal{A}}
{\int_{\mathcal{A}}{A_{\rm{b}}(\mathbf{s},\mathbf{s}^{\prime})}}\nonumber\\
&\times\delta (\mathbf{s}_2-\mathbf{s})\delta (\mathbf{s}^\prime-\mathbf{s}_{2}^{\prime})\mathrm{d}\mathbf{s}\mathrm{d}\mathbf{s}^{\prime}=A_{\rm{b}}(\mathbf{s}_2,\mathbf{s}_{2}^{\prime}).
\end{align}
Moreover, $I_2$ can be calculated as follows:
\begin{align}
I_2&=\int_{\mathcal{A}}{\int_{\mathcal{A}}{}}\Big( \delta (\mathbf{s}_2-\mathbf{s}_1)-\frac{\mu}{1+\mu g_{\mathrm{e}}}h_{\mathrm{e}}(\mathbf{s}_2)h_{\mathrm{e}}^{*}(\mathbf{s}_1) \Big) \delta (\mathbf{s}_1-\mathbf{s}_{1}^{\prime})\notag\\
&\times\Big( \delta (\mathbf{s}_{1}^{\prime}-\mathbf{s}_{2}^{\prime})-\frac{\mu}{1+\mu g_{\mathrm{e}}}h_{\mathrm{e}}(\mathbf{s}_{1}^{\prime})h_{\mathrm{e}}^{*}(\mathbf{s}_{2}^{\prime}) \Big) \mathrm{d}\mathbf{s}_1\mathrm{d}\mathbf{s}_{1}^{\prime}\notag\\
&=\int_{\mathcal{A}}{}\Big( \delta (\mathbf{s}_2-\mathbf{s}_{1}^{\prime})-\frac{\mu}{1+\mu g_{\mathrm{e}}}h_{\mathrm{e}}(\mathbf{s}_2)h_{\mathrm{e}}^{*}(\mathbf{s}_{1}^{\prime}) \Big) \notag\\
&\times\Big( \delta (\mathbf{s}_{1}^{\prime}-\mathbf{s}_{2}^{\prime})-\frac{\mu}{1+\mu g_{\mathrm{e}}}h_{\mathrm{e}}(\mathbf{s}_{1}^{\prime})h_{\mathrm{e}}^{*}(\mathbf{s}_{2}^{\prime}) \Big) \mathrm{d}\mathbf{s}_{1}^{\prime}\notag\\
&=\delta (\mathbf{s}_2-\mathbf{s}_{2}^{\prime})+\frac{\mu}{1+\mu g_{\mathrm{e}}}\left( \frac{\mu g_{\mathrm{e}}}{1+\mu g_{\mathrm{e}}}-2 \right) h_{\mathrm{e}}(\mathbf{s}_2)h_{\mathrm{e}}^{*}(\mathbf{s}_{2}^{\prime})\notag.
\end{align}
Recalling that $\mu =-\frac{1}{g_{\mathrm{e}}}\pm \frac{1}{g_{\mathrm{e}}\sqrt{1+\overline{\gamma }_{\mathrm{e}}g_{\mathrm{e}}}}$ yields
\begin{align}
\frac{\mu}{1+\mu g_{\mathrm{e}}}\left( \frac{\mu g_{\mathrm{e}}}{1+\mu g_{\mathrm{e}}}-2 \right) =\frac{\mu(-2-\mu g_{\mathrm{e}})}{(1+\mu g_{\mathrm{e}})^2}=\overline{\gamma }_{\mathrm{e}},
\end{align}
which leads to
\begin{align}\label{I2}
I_2=\delta (\mathbf{s}_2-\mathbf{s}_{2}^{\prime})+\overline{\gamma }_{\mathrm{e}}h_{\mathrm{e}}(\mathbf{s}_2)h_{\mathrm{e}}^{*}(\mathbf{s}_{2}^{\prime})
=A_{\rm{e}}(\mathbf{s}_2-\mathbf{s}_{2}^{\prime}).
\end{align}
By inserting \eqref{I1} and \eqref{I2} into \eqref{Eigenvalue_Equation_Basic2}, the results of \eqref{C_derivation} follow immediately.
\subsection{Proof of \textbf{Lemma \ref{lem_eigenvalue}}}\label{Appendix:D}
Upon observing \eqref{C_derivation}, the Hermitian operator $\breve{C}(\mathbf{s}_2,\mathbf{s}_{2}^{\prime})\triangleq\overline{\gamma }_{\mathrm{b}}h_{\mathrm{b}}(\mathbf{s}_2)h_{\mathrm{b}}^{*}(\mathbf{s}_{2}^{\prime})-\lambda _B \overline{\gamma }_{\mathrm{e}} h_{\mathrm{e}}(\mathbf{s}_2)h_{\mathrm{e}}^{*}(\mathbf{s}_{2}^{\prime})$ is a function of $h_{\mathrm{b}}(\cdot)$ and $h_{\mathrm{e}}(\cdot)$. Therefore, it can be concluded that the eigenfunction of $\breve{C}(\mathbf{s}_2,\mathbf{s}_{2}^{\prime})$ must lie in the subspace spanned by $h_{\mathrm{b}}(\cdot)$ and $h_{\mathrm{e}}(\cdot)$. This means that the eigenvalues and eigenfunctions of $\breve{C}(\mathbf{s}_2,\mathbf{s}_{2}^{\prime})$ can be determined from the following equation:
\setlength\abovedisplayskip{5pt}
\setlength\belowdisplayskip{5pt}
	\begin{equation}\label{D1}
		\begin{split}
			&\int_{\mathcal{A}}\breve{C}(\mathbf{s}_2,\mathbf{s}_{2}^{\prime})(ah^*_{\mathrm{b}}(\mathbf{s}_2)-bh^*_{\mathrm{e}}(\mathbf{s}_2))\mathrm{d}\mathbf{s}_2\\
			&=\xi(ah^*_{\mathrm{b}}(\mathbf{s}'_2)-bh^*_{\mathrm{e}}(\mathbf{s}'_2)),
		\end{split}
	\end{equation}
where $\xi$ denotes the eigenvalue of $\breve{C}(\mathbf{s}_2,\mathbf{s}_{2}^{\prime})$. 

By performing some mathematical manipulations to the left-hand side of \eqref{D1}, we can transform \eqref{D1} as follows:
\begin{equation}\label{Eigenvalue_Equation_Transform1}
		\begin{split}
			&\left( a\overline{\gamma }_{\mathrm{b}}g_{\mathrm{b}}\!-\!b\overline{\gamma }_{\mathrm{b}}\rho \right)\! h^*_{\mathrm{b}}(\mathbf{s}'_2)\!-\!( a\lambda _B  \overline{\gamma }_{\mathrm{e}}\rho ^*\!-\!b\lambda _B  \overline{\gamma }_{\mathrm{e}}g_{\mathrm{e}} ) h^*_{\mathrm{e}}(\mathbf{s}'_2)\\
			&=a\xi h^*_{\mathrm{b}}(\mathbf{s}'_2)-b\xi h^*_{\mathrm{e}}(\mathbf{s}'_2).
		\end{split}			
	\end{equation}
Based on \eqref{Eigenvalue_Equation_Transform1}, we have
	\begin{equation}
			a\xi = a\overline{\gamma }_{\mathrm{b}}g_{\mathrm{b}}-b\overline{\gamma }_{\mathrm{b}}\rho,\
			b\xi=a\lambda _B \overline{\gamma }_{\mathrm{e}}\rho ^*-b\lambda _B \overline{\gamma }_{\mathrm{e}}g_{\mathrm{e}}.		
	\end{equation}
It follows that
	\begin{equation}\label{b_a}
		\frac{\xi-\overline{\gamma}_{\mathrm{b}}g_{\mathrm{b}}}{\overline{\gamma}_{\mathrm{b}}\rho}=-\frac{b}{a},\quad
		\frac{\xi+\lambda _B \overline{\gamma}_{\mathrm{e}}g_{\mathrm{e}}}{\lambda _B \overline{\gamma}_{\mathrm{e}}\rho^*}=\frac{a}{b},
	\end{equation}
which leads to the following equation with respect to $\xi$:
	\begin{equation}
		\frac{\xi-\overline{\gamma}_{\mathrm{b}}g_{\mathrm{b}}}{\overline{\gamma}_{\mathrm{b}}\rho}
		\frac{\xi+\lambda _B \overline{\gamma}_{\mathrm{e}}g_{\mathrm{e}}}{\lambda _B \overline{\gamma}_{\mathrm{e}}\rho^*}+1=0,
	\end{equation}
or equivalently,
\begin{equation}
	\xi^2-(\overline{\gamma}_{\mathrm{b}}g_{\mathrm{b}}-\lambda _B \overline{\gamma}_{\mathrm{e}}g_{\mathrm{e}})\xi
	-\lambda _B \overline{\gamma}_{\mathrm{b}}\overline{\gamma}_{\mathrm{e}}g_{\mathrm{b}}g_{\mathrm{e}}(1-\overline{\rho})
	=0.
\end{equation}
The solutions to the above equation are given by $\xi=\xi_1$ and $\xi=\xi_2$, where $\xi_1$ and $\xi_2$ are given in \eqref{eigenvalue_C_solution1} and \eqref{eigenvalue_C_solution2}, respectively. In other words, $\xi=\xi_1$ and $\xi=\xi_2$ are the non-zero eigenvalues of $\breve{C}(\mathbf{s}_2,\mathbf{s}_{2}^{\prime})$. 

The above arguments imply that the eigen-decomposition of $\breve{C}(\mathbf{s}_2,\mathbf{s}_{2}^{\prime})$ can be written as follows:
\begin{align}
\breve{C}(\mathbf{s}_2,\mathbf{s}_{2}^{\prime})=\sum\nolimits_{i=1}^{2}\xi_i\varphi_i(\mathbf{s}_2)\varphi_i^{*}(\mathbf{s}_{2}^{\prime}),
\end{align}
where $\{\varphi_i(\cdot)\}_{i=1}^{2}$ represents an orthonormal basis on ${\mathcal{A}}$. Furthermore, it holds that 
\begin{align}
\delta (\mathbf{s}_2-\mathbf{s}_{2}^{\prime})=\sum\nolimits_{i=1}^{\infty}\varphi_i(\mathbf{s}_2)\varphi_i^{*}(\mathbf{s}_{2}^{\prime}),
\end{align}
where $\{\varphi_i(\cdot)\}_{i=1}^{\infty}$ represents a complete orthonormal basis on ${\mathcal{A}}$. Taken together, we have
\begin{equation}
\begin{split}
C(\mathbf{s}_2,\mathbf{s}_2^{\prime})&=\breve{C}(\mathbf{s}_2,\mathbf{s}_{2}^{\prime})+(1-\lambda _B)\delta (\mathbf{s}_2-\mathbf{s}_{2}^{\prime})\\
&=\sum\nolimits_{i=1}^{2}(\xi_i+1-\lambda _B)\varphi_i(\mathbf{s}_2)\varphi_i^{*}(\mathbf{s}_{2}^{\prime})\\
&+\sum\nolimits_{i=3}^{\infty}(1-\lambda _B)\varphi_i(\mathbf{s}_2)\varphi_i^{*}(\mathbf{s}_{2}^{\prime}),
\end{split}
\end{equation}
which represents the eigen-decomposition of $C(\mathbf{s}_2,\mathbf{s}_2^{\prime})$. As a result, the eigenvalues of $C(\mathbf{s}_2,\mathbf{s}_2^{\prime})$ are given by
\begin{equation}
	\begin{split}
		&\lambda_{C,1}=\xi_1-\lambda_B+1,\ \lambda_{C,2}=\xi_2-\lambda_B+1,\\
		&\lambda_{C,3}=\ldots=\lambda_{C,\infty}=-\lambda_B+1,
	\end{split}
\end{equation}
which completes the proof of \textbf{Lemma \ref{lem_eigenvalue}}.
\vspace{-5pt}
\subsection{Proof of \textbf{Theorem \ref{lem_principal}}}\label{Appendix:E}
It follows from \eqref{equation_trans3} that $\lambda_B=1$ or $\lambda_B=\xi_1+1$ or $\lambda_B=\xi_2+1$. For $\lambda_B=\xi_1+1$ or $\lambda_B=\xi_2+1$, we have
\begin{align}
		\lambda_B=\frac{\Delta\pm
			\sqrt{\Delta^2+4\lambda _B \overline{\gamma}_{\mathrm{b}}\overline{\gamma}_{\mathrm{e}}g_{\mathrm{b}}g_{\mathrm{e}}(1-\overline{\rho})}}{2}.
\end{align} 
This leads to $(2\lambda _B-\Delta )^2=\Delta ^2+4\lambda _B  \overline{\gamma }_{\mathrm{b}}\overline{\gamma }_{\mathrm{e}}g_{\mathrm{b}}g_{\mathrm{e}}(1-\overline{\rho })$, which can be further simplified as follows:
	\begin{equation}
		\left( 1+\overline{\gamma }_{\mathrm{e}}g_{\mathrm{e}} \right) (\lambda _{B}-1)^{2}-\Xi (\lambda _B-1)-\overline{\gamma }_{\mathrm{b}}\overline{\gamma }_{\mathrm{e}}g_{\mathrm{b}}g_{\mathrm{e}}(1-\overline{\rho })=0.
	\end{equation}
The solutions to the above equation are given by
\begin{equation}
	\lambda _{B,1}=1+\frac{\Xi + \sqrt{\Xi ^2+\Gamma}}{2\left( 1+\overline{\gamma }_{\mathrm{e}}g_{\mathrm{e}} \right)},\ \lambda _{B,2}=1+\frac{\Xi - \sqrt{\Xi ^2+\Gamma}}{2\left( 1+\overline{\gamma }_{\mathrm{e}}g_{\mathrm{e}} \right)},\nonumber
\end{equation}
where $\Xi =\overline{\gamma }_{\mathrm{b}}g_{\mathrm{b}}-\overline{\gamma }_{\mathrm{e}}g_{\mathrm{e}}+\overline{\gamma }_{\mathrm{b}}\overline{\gamma }_{\mathrm{e}}g_{\mathrm{b}}g_{\mathrm{e}}(1-\overline{\rho })$, and $\Gamma =4\left( 1+\overline{\gamma }_{\mathrm{e}}g_{\mathrm{e}} \right) \overline{\gamma }_{\mathrm{b}}\overline{\gamma }_{\mathrm{e}}g_{\mathrm{b}}g_{\mathrm{e}}(1-\overline{\rho })$. Since $\lambda _{B,1}\geq0\geq \lambda _{B,2}$, the principal eigenvalue of $B(\mathbf{s}_1,\mathbf{s}_1^{\prime})$ is $\lambda _{B,1}$, which can be further simplified as follows:
{\setlength\abovedisplayskip{1pt}
\setlength\belowdisplayskip{2pt}
\begin{align}
	\lambda _{B}^{\max}=\lambda _{B,1}\!&=1+\frac{\Xi +\!\sqrt{\left( \overline{\gamma }_{\mathrm{b}}g_{\mathrm{b}}\left( 1+\overline{\gamma }_{\mathrm{e}}g_{\mathrm{e}}\left( 1\!-\!\overline{\rho } \right) \right) +\overline{\gamma }_{\mathrm{e}}g_{\mathrm{e}} \right) ^2}}{2\left( 1+\overline{\gamma }_{\mathrm{e}}g_{\mathrm{e}} \right)}\nonumber\\
	&=1+\frac{\overline{\gamma }_{\mathrm{b}}g_{\mathrm{b}}\left( 1+\overline{\gamma }_{\mathrm{e}}g_{\mathrm{e}}\left( 1-\overline{\rho } \right) \right)}{1+\overline{\gamma }_{\mathrm{e}}g_{\mathrm{e}}}.
\end{align}}
This completes the proof of \textbf{Theorem \ref{lem_principal}}.

\subsection{Proof of \textbf{Lemma \ref{lem_optimal_u}}}\label{Appendix:F}
Given that the optimal solution to problem \eqref{CAP_MISOSE_Problem_5} corresponds to the principal eigenfunction of $B(\mathbf{s}_1,\mathbf{s}_1^{\prime})$, and noting that $B(\mathbf{s}_1,\mathbf{s}_1^{\prime})$ shares the same eigenfunctions as ${\hat{B}}(\mathbf{s}_1,\mathbf{s}_1^{\prime})=B(\mathbf{s}_1,\mathbf{s}_1^{\prime})-\lambda_B\delta(\mathbf{s}_1-\mathbf{s}_1^{\prime})$, problem \eqref{CAP_MISOSE_Problem_5} can be reformulated as follows:
\begin{equation}\label{CAP_MISOSE_Problem_6}
	\max_{\nu (\mathbf{s}_1)} \int_{\mathcal{A}}{\int_{\mathcal{A}}{}}\nu (\mathbf{s}_1){\hat{B}}(\mathbf{s}_1,\mathbf{s}_1^{\prime})\nu ^*(\mathbf{s}_{1}^{\prime})\mathrm{d}\mathbf{s}_1\mathrm{d}\mathbf{s}_{1}^{\prime}.
\end{equation}
Recalling the definition given in \eqref{C_definition} and the invertible relationship given in \eqref{Q_Inversion}, we obtain
\begin{equation} \!\!{\hat{B}}(\mathbf{s}_1,\mathbf{s}_1^{\prime})\!=\!\int_{\mathcal{A}}{\int_{\mathcal{A}}}Q(\mathbf{s}_1,\mathbf{s}_2)C(\mathbf{s}_2,\mathbf{s}_{2}^{\prime})Q(\mathbf{s}_{2}^{\prime},\mathbf{s}_{1}^{\prime})\mathrm{d}\mathbf{s}_2\mathrm{d}\mathbf{s}_{2}^{\prime}.
\end{equation}
By substituting the above expression into \eqref{CAP_MISOSE_Problem_6} and using the fact that $Q(\mathbf{s}_{2}^{\prime},\mathbf{s}_{1}^{\prime})=Q^*(\mathbf{s}_{1}^{\prime},\mathbf{s}_{2}^{\prime})$, the objective function of \eqref{CAP_MISOSE_Problem_6} can be rewritten as follows:
\begin{align}
	&\int_{\mathcal{A}}{\int_{\mathcal{A}}{}}\nu (\mathbf{s}_1){\hat{B}}(\mathbf{s}_1,\mathbf{s}_1^{\prime})\nu ^*(\mathbf{s}_{1}^{\prime})\mathrm{d}\mathbf{s}_1\mathrm{d}\mathbf{s}_{1}^{\prime}\notag\\
	&=\int_{\mathcal{A}}{\int_{\mathcal{A}}{}}C(\mathbf{s}_2,\mathbf{s}_{2}^{\prime})\int_{\mathcal{A}}{}\nu (\mathbf{s}_1)Q(\mathbf{s}_1,\mathbf{s}_2)\mathrm{d}\mathbf{s}_1\nonumber\\
	&~\times\int_{\mathcal{A}}{}\nu^*(\mathbf{s}_{1}^{\prime})Q^*(\mathbf{s}_{1}^{\prime},\mathbf{s}_{2}^{\prime})\mathrm{d}\mathbf{s}_{1}^{\prime}\mathrm{d}\mathbf{s}_2\mathrm{d}\mathbf{s}_{2}^{\prime}\nonumber\\
	&=\int_{\mathcal{A}}{\int_{\mathcal{A}}{}}u\left( \mathbf{s}_2 \right) C(\mathbf{s}_2,\mathbf{s}_{2}^{\prime})u^*\left( \mathbf{s}_{2}^{\prime} \right) \mathrm{d}\mathbf{s}_2\mathrm{d}\mathbf{s}_{2}^{\prime},
\end{align}
where $u(\mathbf{s})=\int_{\mathcal{A}}{\nu }(\mathbf{s}_1)Q\left({\mathbf{s}}_1,\mathbf{s}\right){\rm{d}}{\mathbf{s}}_1$. Therefore, we can obtain the optimal $\nu\left( \mathbf{s} \right)$ through solving the following problem:
\begin{equation}\label{Optimal_Rate_Current_Medium_Step}
u^{\star}({\mathbf{s}})=	\argmax_{u(\mathbf{s})} \int_{\mathcal{A}}{\int_{\mathcal{A}}{}}u\left( \mathbf{s} \right) C(\mathbf{s},\mathbf{s}^{\prime})u^*\left( \mathbf{s}^{\prime} \right) \mathrm{d}\mathbf{s}\mathrm{d}\mathbf{s}^{\prime},
\end{equation}
and then performing the transformation $\nu^{\star}(\mathbf{s}_1)=\int_{{\mathcal{A}}}u^{\star}({\mathbf{s}})\hat{Q}\left( \mathbf{s},\mathbf{s}_1 \right){\rm{d}}{\mathbf{s}}$. It can be observed from \eqref{Optimal_Rate_Current_Medium_Step} that the optimal $u(\mathbf{s})$ is aligned with the principal eigenfunction of $C(\mathbf{s},\mathbf{s}^{\prime})$. The final results follow immediately.

\subsection{Proof of \textbf{Lemma \ref{lem_planar_capa}}}\label{Appendix:G}
With the planar CAPA, the channel gain for user $k$ can be written as follows:
\begin{equation}\label{G1}
\begin{split}
&	\!\!g_k\!=\!\int_{-\frac{L_z}{2}}^{\frac{L_z}{2}}{\int_{-\frac{L_x}{2}}^{\frac{L_x}{2}}{h_{k}^{*}(x,z)h_k(x,z)\mathrm{d}x}\mathrm{d}z}
	=\frac{k_{0}^{2}\eta ^2r_k\Psi _k}{4\pi}\\
&\!\!\times\underset{I_3}{\underbrace{\int_{-\frac{L_z}{2}}^{\frac{L_z}{2}}\int_{-\frac{L_x}{2}}^{\frac{L_x}{2}}
		\!\frac{\mathrm{d}x\mathrm{d}z}{(x^2+z^2-2r_k\left( \Phi _kx+\Theta _kz \right) +r_{k}^{2})^{\frac{3}{2}}}}},
\end{split}
\end{equation}
To compute the double integral $I_1$, we first use \cite[Eq. (2.264.5)]{integral} to calculate the inner integral with respect to $x$, which yields
\begin{align}\label{I3}
	I_3&=\int_{-\frac{L_z}{2}}^{\frac{L_z}{2}}{\frac{\frac{\frac{L_x}{2}-r_k\Phi _k}{\sqrt{\left( z-r_k\Theta _k \right) ^2+r_{k}^{2}\Psi _{k}^{2}+\left( \frac{L_x}{2}-r_k\Phi _k \right) ^2}}}{\left( z-r_k\Theta _k \right) ^2+r_{k}^{2}\Psi _{k}^{2}}\mathrm{d}z}\notag\\
	&+\int_{-\frac{L_z}{2}}^{\frac{L_z}{2}}{\frac{\frac{\frac{L_x}{2}+r_k\Phi _k}{\sqrt{\left( z-r_k\Theta _k \right) ^2+r_{k}^{2}\Psi _{k}^{2}+\left( \frac{L_x}{2}+r_k\Phi _k \right) ^2}}}{\left( z-r_k\Theta _k \right) ^2+r_{k}^{2}\Psi _{k}^{2}}\mathrm{d}z}\notag\\
	&=\int_{-\frac{L_z}{2}-r_k\Theta _k}^{\frac{L_z}{2}-r_k\Theta _k}{\frac{\frac{L_x}{2}-r_k\Phi _k}{\left( t^2\!+\!r_{k}^{2}\Psi _{k}^{2} \right) \sqrt{t^2\!+\!r_{k}^{2}\Psi _{k}^{2}\!+\!\left( \frac{L_x}{2}\!-\!r_k\Phi _k \right) ^2}}\mathrm{d}t}\notag\\
	&+\int_{-\frac{L_z}{2}-r_k\Theta _k}^{\frac{L_z}{2}-r_k\Theta _k}{\frac{\frac{L_x}{2}+r_k\Phi _k}{\left( t^2\!+\!r_{k}^{2}\Psi _{k}^{2} \right) \sqrt{t^2\!+\!r_{k}^{2}\Psi _{k}^{2}\!+\!\left( \frac{L_x}{2}\!+\!r_k\Phi _k \right) ^2}}\mathrm{d}t}.
\end{align}
Both integrals with respect to $t$ in \eqref{I3} can be evaluated with the aid of \cite[Eq. (2.284)]{integral}, which gives
\begin{align}\label{I3_2}
	I_3&=\frac{1}{r_k\Psi _k}\times\notag\\
	&\left[ \arctan \left( \frac{\left( \frac{L_x}{2r_k}-\Phi _k \right) \left( \frac{L_z}{2r_k}-\Theta _k \right)}{\Psi _k\sqrt{\left( \frac{L_z}{2r_k}-\Theta _k \right) ^2+\Psi _{k}^{2}+\left( \frac{L_x}{2r_k}-\Phi _k \right) ^2}} \right)  \right.\notag\\
	&+\arctan \left( \frac{\left( \frac{L_x}{2r_k}-\Phi _k \right) \left( \frac{L_z}{2r_k}+\Theta _k \right)}{\Psi _k\sqrt{\left( \frac{L_z}{2r_k}+\Theta _k \right) ^2+\Psi _{k}^{2}+\left( \frac{L_x}{2r_k}-\Phi _k \right) ^2}} \right) \notag\\
	&+\arctan \left( \frac{\left( \frac{L_x}{2r_k}+\Phi _k \right) \left( \frac{L_z}{2r_k}-\Theta _k \right)}{\Psi _k\sqrt{\left( \frac{L_z}{2r_k}-\Theta _k \right) ^2+\Psi _{k}^{2}+\left( \frac{L_x}{2r_k}+\Phi _k \right) ^2}} \right) \notag\\
	&\left.+\arctan \left( \frac{\left( \frac{L_x}{2r_k}+\Phi _k \right) \left( \frac{L_z}{2r_k}+\Theta _k \right)}{\Psi _k\sqrt{\left( \frac{L_z}{2r_k}+\Theta _k \right) ^2+\Psi _{k}^{2}+\left( \frac{L_x}{2r_k}+\Phi _k \right) ^2}} \right) \right] \notag\\
	&=\frac{1}{r_k\Psi _k}\!\sum_{x\in \mathcal{X} _k}{\sum_{z\in \mathcal{Z} _k}\!{\arctan \!\Bigg( \frac{xz}{\Psi _k\sqrt{\Psi _{k}^{2}+x^2+z^2}} \Bigg)}}.
\end{align}
Consequently, substituting \eqref{I3_2} into \eqref{G1}, we obtain the results of \eqref{g_planar_capa}.

We next move to the channel correlation (inner product), which can be expressed as follows:
\begin{equation}
	\rho =\int_{-\frac{L_z}{2}}^{\frac{L_z}{2}}{\int_{-\frac{L_x}{2}}^{\frac{L_x}{2}}{h_{\mathrm{b}}^{*}(x,z)h_{\mathrm{e}}(x,z)\mathrm{d}x}\mathrm{d}z}.
\end{equation}
We can evaluate the above integrals based on the Chebyshev-Gauss quadrature rule \cite{math}, i.e., 
\begin{equation}
	\int_a^b{f\left( x \right) \mathrm{d}}x\approx \frac{b-a}{2}\sum_{t=1}^T{\frac{\pi}{T}\sqrt{1-\psi _{t}^{2}}f\left( \frac{b-a}{2}\psi _t+\frac{b+a}{2} \right)},
\end{equation}
which yields the results presented in \eqref{rho_planar_capa}. Note that the value of $T$ should be carefully chosen to balance computational complexity and numerical accuracy.

\subsection{Proof of \textbf{Lemma \ref{lem_planar_spda}}}\label{Appendix:I}
By defining $\epsilon_k=\frac{d}{r_k} \ll 1$, we can rewrite \eqref{g_planar_spda0} as follows:
\begin{equation}\label{i1}
	g_{k}^{\mathrm{s}}=\frac{Ak_{0}^{2}\eta ^2\Psi _k}{4\pi r_{k}^{2}}\sum_{m_x\in \mathcal{M} _x}{\sum_{m_z\in \mathcal{M} _z}{f_{\mathrm{s}}}}\left( m_x\epsilon _k,m_z\epsilon _k \right),
\end{equation}
Here, $f_{\mathrm{s}}\left( x,z \right) \triangleq (x^2+z^2-2\Phi_kx-2\Omega_kz+1)^{-\frac{3}{2}}$ is a function defined over the square region ${\mathcal{S}}_k\triangleq\{ (x,z) \mid -\frac{M_x\epsilon _k}{2}\leq x\leq \frac{M_x\epsilon _k}{2},-\frac{M_z\epsilon _k}{2}\leq z\leq \frac{M_z\epsilon _k}{2} \}  $, which is divided into $M_xM_z$ sub-squares, each with an area of $\epsilon _k^2$. Since $\epsilon _k\ll 1$, it follows that $f_{\mathrm{s}}\left( x,z \right) \approx f_{\mathrm{s}}\left( m_x\epsilon _k,m_z\epsilon _k \right) $ for $\forall \left( x,z \right) \in \left\{ \left( x,z \right) \mid \left( m_x-\frac{1}{2} \right) \epsilon _k\leq x\leq \left( m_x+\frac{1}{2} \right) \epsilon _k,\left( m_z-\frac{1}{2} \right) \epsilon _k\leq z \right.$ $\left.\leq \left( m_z+\frac{1}{2} \right) \epsilon _k \right\}$. By the concept of double integral, it has
\begin{equation}
	\sum_{m_x\in \mathcal{M} _x}\sum_{m_z\in \mathcal{M} _z}\!{f_{\mathrm{s}}\left( m_x\epsilon _k,m_z\epsilon _k \right) \epsilon _k^2}\approx \iint_{{\mathcal{S}}_k}{f_{\mathrm{s}}\left( x,z \right) \mathrm{d}x\mathrm{d}z}.  
\end{equation}
Therefore, \eqref{i1} can be rewritten as follows:
\begin{equation}
	g_{k}^{\mathrm{s}}\approx \frac{\zeta _{\mathrm{oc}}k_{0}^{2}\eta ^2\Psi _k}{4\pi}\int_{-\frac{M_z\epsilon _k}{2}}^{\frac{M_z\epsilon _k}{2}}{\int_{-\frac{M_x\epsilon _k}{2}}^{\frac{M_x\epsilon _k}{2}}{}}f_{\mathrm{s}}\left( x,z \right) \mathrm{d}x\mathrm{d}z, 
\end{equation}
which can be calculated with the aid of \cite[Eqs. (2.264.5) \& (2.284)]{integral}. The final results follow immediately.
\end{appendix}

\bibliographystyle{IEEEtran}
\bibliography{mybib}
\end{document}